\documentclass[12pt]{article}
\pdfoutput=1
\newcommand{\bxi}{\boldsymbol{\xi}}
\DeclareMathAlphabet{\bi}{OML}{cmm}{b}{it}

\usepackage{authblk} % author and affiliation blocks
\usepackage{amsmath}
\usepackage{amssymb}
\usepackage{amsthm, amsfonts}
\usepackage{mathrsfs, mathtools}
\usepackage{makecell}

\usepackage[table,xcdraw]{xcolor} % For coloring
\usepackage{booktabs}

\usepackage{siunitx} % Used for SI units
\usepackage{graphicx} % For figures
\usepackage{float}

\usepackage{subcaption}
\usepackage{setspace}
\usepackage[a4paper, margin=1in]{geometry}

\usepackage[comma]{natbib}

\usepackage{url}

\newcommand{\mycomment}[1]{}

\title{\LARGE{\textbf{Direct probabilistic IMPT treatment planning with setup and range errors for neuro-oncological patients}}}

\author{Jelte R. de Jong$^1$, Sebastiaan Breedveld$^3$,
Steven J. M. Habraken$^{2,4}$, Mischa S. Hoogeman$^{2,3}$, Jenneke I. de Jong$^{2,3}$, Danny Lathouwers$^1$ and Zoltán Perkó}

\affil[1]{\footnotesize{Delft University of Technology, Department of Radiation, Science and Technology, Delft, The Netherlands}}
\affil[2]{\footnotesize{Holland Proton Therapy Centre Delft, Delft, the Netherlands}}
\affil[3]{\footnotesize{Erasmus MC Cancer Institute, University Medical Center Rotterdam, Department of Radiotherapy, Rotterdam, the Netherlands}}
\affil[4]{\footnotesize{Leiden University Medical Center, Department of Radiation Oncology, Leiden, the Netherlands}}

\date{\small{\today}}

\begin{document}

\bibliographystyle{dcu}
\def\newblock{\ }%
\setcitestyle{authoryear,open={(},close={)}}

\maketitle

\begin{abstract}

\textit{Objective}. To show clinical feasibility of a previously proposed probabilistic planning approach that can precisely optimize for clinical goals with patient-specific acceptance probabilities on a neuro-oncological patient group, we compared probabilistic plans with (automated) robust plans for one patient (group A) that could achieve sufficient clinical target coverage and for four patients (group B) where target coverage had to be compromised due to organ-at-risk (OAR) dose constraints. 
\textit{Approach}. The probabilistic approach is percentile-based and uses the fact that a (dose) percentile can be approximated as a linear combination of its expected value and standard deviation. The optimization has a nested structure: the inner optimization optimizes the beam weights for a given percentile estimate, while an outer loop iteratively updates and improves the accuracy of the percentile estimate. For every outer iteration, the optimization is warm-started from the previous iteration. Percentiles are efficiently calculated by sampling a polynomial chaos expansion of the dose-influence matrix.
\textit{Main results}. The patient in group A achieved cumulative OAR dose reductions (of OAR-related DVH-metrics) of \qty{19}{\gray}RBE, for identical target coverage. Target coverage improved for all patients in group B (the 10th percentile of $D_{99.8\%}$ increased up to \qty{0.93}{\gray}RBE), at the same time reaching cumulative OAR dose reductions (of OAR-related DVH-metrics) up to \qty{33}{\gray}RBE. Probabilistic plans were optimized in \qtyrange{44}{141}{\hour}. For two representative patients, eliminating warm-starting (i.e., the outer loop) from the approach reduced total optimization times to below \qty{10}{\hour} (which took originally \qty{80}{\hour} and \qty{141}{\hour}).
\textit{Significance}. Compared to robust optimization methods, the probabilistic approach achieves improved trade-offs between probabilistic target coverage and OAR sparing, potentially leading to better treatments.

\end{abstract}

%\vspace{2pc}

\section{Introduction} \label{sec:Introduction}

In particle therapies such as proton therapy, treatment uncertainties often lead to differences between the planned and delivered dose distributions. Uncertainties in patient setup, stopping power prediction (i.e., proton range errors), anatomical changes or uncertainties in clinical target volume (CTV) and organ-at-risk (OAR) delineation could potentially lead to insufficient target coverage or unnecessarily high dose to OARs if they are not properly accounted for \citep{Schaffner1998,Lomax2008}. Especially for treatment modalities where steep dose gradients are common, (e.g., intensity-modulated proton therapy - IMPT), it is essential to control the plan robustness by including uncertainties in the plan optimization \citep{Lomax2008_1}. \newline

The clinical standard for photon therapy is to extend the dose margin beyond the CTV into the planning target volume (PTV), which works well for modalities where the static dose cloud approximation holds \citep{Bortfeld2004}. For IMPT however, dose distributions tend to deform significantly under uncertainty, which is why scenario-based optimization is the current clinical standard. Mini-max robust optimization is often used, where the worst-case across a scenario set is optimized. Plan robustness depends both on the type of mini-max approach \citep{Pflugfelder2008,Chen2012,vanDijk2016,Janson2024} and on the choice of the scenario set: it may become overly conservative if a few scenarios dominate, or be limited if the scenario set was chosen too narrow \citep{vanderVoort2016,Zhang2021}. Although robust optimization methods show superior results compared to PTV-based approaches for IMPT \citep{Kiselev2025,Liu2012,Liu2012_2,Liu2013,vanDijk2016}, mini-max robust optimization optimizes for the worst-case outcome across the scenario set, and does not account for the probability of occurrence of these scenarios. As a result, a (rare) extreme scenario contributes equally to the objective as a near-nominal scenario, whereas in reality large setup and range errors are much less probable to occur. Therefore, mini-max robust optimization cannot directly optimize for probabilistic goals. \newline

Robust optimization is usually accompanied by robust evaluation, which in the Netherlands is based on voxel-wise minimum (VWmin) and maximum (VWmax) metrics across included scenarios \citep{Korevaar2019}. In clinical practice, these metrics tend to lead to conservative target coverage and to inter-patient and inter-center variation \citep{RojoSantiago2021,RojoSantiago2023,RojoSantiago2024}. In general, it is not trivial how to choose amongst the many options of worst-case evaluation methods and metrics that are available, as these are mathematically and statistically inconsistent \citep{Sterpin2021}. The lack of consensus on how to evaluate plan robustness leads to significant variation between clinical centers \citep{Kaplan2022}. This calls for the development of different optimization and evaluation methods, that capture the statistical nature of uncertainties in a more consistent way rather than using a discrete scenario set.

Probabilistic planning approaches are promising for this purpose \citep{Sterpin2024}, as they model uncertainties as continuous distributions, which is more realistic than assuming that only discrete scenarios can occur. Whilst probabilistic approaches can be approximated by weighting discrete scenarios, as in stochastic robust optimization \citep{Unkelbach2018}, this would still require careful selection of the scenario set. To avoid this, indirect probabilistic approaches have been proposed, where the robustness settings of a mini-max optimization are tuned to match probabilistic target coverage or OAR dose, e.g., such that CTV coverage is achieved in 90\% of the scenarios. One such approach is the development of \textit{robustness recipes}, that allow to select the robustness settings corresponding to the given setup and range uncertainties \citep{vanderVoort2016}. More recently, a probabilistic re-optimization approach was introduced \citep{deJong2025}, where robustness settings are adjusted based on comprehensive probabilistic evaluation. Resulting plans showed improved trade-offs between target coverage and OAR sparing. \newline

Direct probabilistic optimization approaches have also been proposed, ranging from expected dose and variance-based methods \citep{Fabiano2022,Chu2005, Cristoforetti2025,Fredriksson2012,Wahl2018} to percentile-based approaches in photon therapy \citep{Tilly2019,Mescher2017,Gordon2010}. However, percentile-based optimization remains challenging for two main reasons: a) probability density functions (PDFs) of quantities of interest (e.g., voxel dose) are complex in practice and are not guaranteed to be close to Gaussian distributed, and b) sufficient error scenarios need to be calculated to accurately estimate the desired percentiles.

Regarding reason (a), various studies have simplified the optimization by assuming Gaussian-distributed dose metrics (e.g., voxel dose or biological effective dose), showing that probabilistic goals could be steered into the desired direction \citep{Sobotta2010,Fabiano2022}. Also different percentile proxies have been introduced: \citet{Tilly2019} optimized for probabilistic DVH-constraints based on iteratively tuning a conditional value-at-risk (CVaR) based metric in photon planning, while \citet{Fredriksson2026} proposes to use a percentile estimate based on a trimmed power mean function. While these proxies provide an efficient way for percentile estimation, it is not guaranteed that the exact percentile is actually optimized for. 

In photon therapy, sampling of error scenarios (regarding reason (b)) can be done more efficiently using the static dose cloud approximation \citep{Gordon2010,Mescher2017,Bohoslavsky2013,Tilly2019}. However, as this assumption is generally not valid for proton therapy, alternative fast sampling techniques have been developed that do not rely on a static dose cloud \citep{JennekedeJong2026,Vazquez2023}. In this work we use polynomial chaos expansion (PCE) \citep{Perk2016} for uncertainty quantification (of continuous setup and range errors) and efficient sampling of error scenarios. \newline 

With all mentioned approaches, it remains difficult to optimize for probabilistic goals, because the objectives and constraints depend on continuous distributions that were approximated in different aspects, namely by distribution shape, percentile surrogates and limited number of samples. Therefore, in previous work \citep{JRdeJong2026}, we generalized the approach by \citet{Fabiano2022} - where voxel dose percentiles were estimated by a linear combination of expected value and standard deviation - allowing us to optimize for exact personalized acceptable risks of target under- and overdosage, as well as OAR overdosage. In particular, constructing a PCE of the dose-influence matrix allows us to efficiently sample \num{100000} error scenarios, so that accurate percentiles from non-Gaussian (dose) distributions can be obtained without using the static dose cloud approximation.

The aim of this study is to demonstrate the feasibility of our probabilistic planning approach on a recently treated neuro-oncological patient group. To ensure scalability towards typical optimization problem sizes, a memory-efficient representation of the covariance matrix was introduced. We compare the probabilistic plan results against worst-case robust plans by evaluating a) whether OAR dose can be reduced for patients with adequate target coverage, and b) whether coverage can be improved for patients in whom coverage could not be achieved in the clinical robust plans.

\section{Methods and Materials} \label{sec:Methods}

% Patient numbering convetion:
% patient 4 (robust) -> patient 1
% patient 10 (robust) -> patient 2
% patient 12 (robust) -> patient 3
% patient 32 (robust) -> patient 4
% patient 35 (robust) -> patient 5

\subsection{Patient cohort and treatment setup}
\label{subsec:clinicalPlanSetting}

Our probabilistic planning approach was applied to five neuro-oncological patients, who were treated at the Holland Proton Therapy Center (HollandPTC) in Delft in 2023. All plans were optimized by multi-field optimization (MFO) IMPT using two or three beams. Prescribed doses were \qty{50.4}{\gray}RBE (1/5) and \qty{59.4}{\gray}RBE (4/5), delivered in fractions of \qty{1.8}{\gray}RBE. Plan comparisons were performed for five selected patients from previous work \citep{deJong2025}: one patient (group A: patient 1) whose clinical plan achieved adequate target coverage, and four patients (group B: patients 2 to 5) whose clinical plans could not achieve sufficient target coverage due to limiting OAR constraints.

Dose calculations were performed using the Astroid dose engine \citep{Kooy2010}, which was tuned for IMPT at HollandPTC \citep{Kouwenberg2021} to give similar dose distributions as their clinical results. Couch and gantry angles, as well as range shifter settings were taken from the clinical RayStation plans of previous work \citep{deJong2025}, while the mini-max robust (Section \ref{subsec:minimaxRobust}) and probabilistic (Section \ref{subsec:probApproach}) plans were independently optimized, after which they were scaled (Section \ref{subsec:probWeights}). Therefore, the clinical RayStation plans formed the basis for the beam geometry and optimization goals, but were not used as a reference plan for comparison.

\subsection{Generation of mini-max robust plans}
\label{subsec:minimaxRobust}
Mini-max robust plans were optimized by the in-house developed Erasmus-iCycle, an automated multi-criteria optimization (MCO) software \citep{Breedveld2012}. Using a predefined \textit{wish-list} (see Appendix \ref{app:wishlists}), cost functions were optimized based on priority, subject to hard constraints. The robust plans were optimized using a voxel-wise worst-case mini-max approach using 21 scenarios for a $\pm \qty{3}{\milli\meter}$ setup error and 3\% range error (as in \citet{deJong2025}), based on the clinical protocol. \newline

Robust plan optimization started by optimizing for the highest priority objective first, after which remaining objectives were optimized sequentially in decreasing order of priority. After an objective was optimized, it was reformulated as a slightly relaxed constraint. Spot positions were determined by Erasmus-iCycle using \qty{1}{\milli\meter} lateral spacing and between \num{1.0}\% and \num{1.6}\% energy spacing, depending on the patient. Hammersley sequence sampling (as implemented in Erasmus-iCycle) was used for voxel selection of larger structures, where about \numrange{5000}{10000} voxels were selected. The voxel density used for the target was set to 50 voxels/\unit{cc}. For small structures ($\lesssim \num{5000}$ voxels), all voxels were included. 

Robust plans were first optimized to satisfy \textit{clinical plan evaluation}, which in HollandPTC is based on voxel-wise minimum (VWmin) and voxel-wise maximum (VWmax) dose metrics using 28 scenarios \citep{Korevaar2019}. These error scenarios consisted of 14 setup error directions defined by the faces and vertices of a cube, each with a magnitude of \qty{3}{\milli\meter}, combined with range errors of $\pm 3\%$. Clinical target coverage was considered sufficient if the VWmin $D_{98\%,CTV}\geq95\% d_p$. For the brainstem core, brainstem surface and optic system, the VWmax $D_{\qty{0.03}{cc}}$ was constrained to \qty{54}{\gray}RBE, \qty{60}{\gray}RBE and \qty{55}{\gray}RBE, respectively. 

\subsection{Polynomial Chaos Expansion}
\label{subsec:PCE}
For the efficient sampling of dose distributions, we first construct a PCE of the dose-influence matrix $D_{ij}(\bxi)$, where $i \in 1, \ldots, N_v$ indexes the voxels and $j \in 1, \ldots, N_b$ indexes the beamlets, as
\begin{equation} \label{eq:DijResponses}
    D_{ij}(\bxi) \approx \sum_{k=0}^P R_{ij}^{(k)} \Psi_k(\bxi),
\end{equation}
where $R_{ij}^{(k)} \in \mathbb{R}^{N_v \times N_b}$ is the $k^{th}$ PCE coefficient corresponding to the $k^{th}$ multi-dimensional Hermite polynomial $\Psi_k(\bxi)$. The expansion is based on $P + 1 = 73$ number of basis vectors. The input uncertainties are assumed to be Gaussian distributed systematic setup errors $\boldsymbol{\Sigma}$ (mean: \qty{0.00}{\milli\meter}, SD: \qty{1.2}{\milli\meter}) and range errors $\rho$ (mean: 1.2\%, SD: 1\%), i.e., $\bxi = (\boldsymbol{\Sigma}, \rho)$. The magnitude of the Gaussian uncertainties were chosen based on the \qty{3}{\milli\meter} margin $M$ following the Van Herk margin recipe \citep{vanHerk2000} for systematic errors only, i.e., $M = 2.5 \boldsymbol{\Sigma}$. The $D_{ij}$ responses in Equation \ref{eq:DijResponses} are computed by the Astroid dose engine and the PCE coefficients are computed by projecting the 217 $D_{ij}$ responses onto $\Psi_k(\bxi)$. An accuracy analysis of the PCE is done in Appendix \ref{app:PCEaccuracy}.
% \ref{app:PCEaccuracy}.

Second, a separate PCE of the dose distribution (i.e., the dose $d_i(\bxi)$) is constructed for \textit{probabilistic plan evaluation} as
\begin{equation} \label{eq:diResponses}
    d_{i}(\bxi) = \sum_{j \in \mathbb{B}} D_{ij}(\bxi) x_j \approx \sum_{k=0}^P r_i^{(k)} \Psi_k(\bxi),
\end{equation}
where the PCE coefficients $r_i$ are determined by spectral projection as well. Probabilistic treatment plan evaluation is done by sampling Equation \ref{eq:diResponses} using $N_s = \num{100000}$ samples (for patient 5, $N_s = \num{25000}$ was used).

\subsection{The probabilistic planning approach} \label{subsec:probApproach}
The methodology in this work extends on the probabilistic planning approach of \citet{JRdeJong2026}, which is percentile-based. The $\alpha$-th percentile of the voxel dose $d_i^{\alpha \%}(\bi{x})$ is the dose value for which $(1-\alpha)$\% of the scenarios lead to larger doses. During optimization, we use a percentile estimate that is based on expected dose $\mathbb{E}[d_i(\bi{x},\bxi)]$ and standard deviation of the dose $\text{SD}[d_i(\bi{x},\bxi)]$, as
\begin{equation} \label{eq:percentileEstimate}
    d_i^{\alpha \%}(\bi{x}) = \mathbb{E}[d_i(\bi{x},\bxi)] \pm \delta_i \cdot \text{SD}[d_i(\bi{x},\bxi)],
\end{equation}
where the multiplicative factor $\delta_i \in \mathbb{R}$ (referred to as the $\delta$-factor) quantifies (in units of standard deviation) the distance between the $\alpha$-th percentile of the voxel dose ($d_i^{\alpha \%}(\bi{x})$) and the expected dose. Compared to the initially proposed approach of \citet{JRdeJong2026}, the following improvements have been made.
\begin{itemize}
  \item Probabilistic constraints are included (e.g., $d_i^{\alpha \%}(\bi{x}) \geq \gamma_i, \, \forall i \in S$ for dose threshold $\gamma$ and structure $S$). Probabilistic mean dose can be used as well, but is not used in this work.
  \item By using a reduced (diagonal-only) representation of the covariance matrix, the memory requirement can be reduced from $\mathcal{O}(N_b^2)$ to $\mathcal{O}(N_b)$, where $N_b$ is the number of beamspots (see Section \ref{paragraph:memEfficient_percEstimate}).
  \item Convergence criteria are defined by smoothing \textit{undamped} percentiles instead of the previous approach of damping percentiles. For particular cases, this can lead to significantly less iterations (see Appendix \ref{app:convCriteria} for details). % \ref{app:convCriteria}
\end{itemize}
The voxel set that is used in the probabilistic plan is identical to the one used in the robust plan (and is thus obtained as well by Hammersley sequence sampling). No further (coarser) voxel selection was needed, because the diagonal-only covariance matrix provided sufficient memory reduction. We refer to this sampled voxel set as \textit{active} voxels.

\begin{figure}
    \centering
    \includegraphics[width=0.8\linewidth]{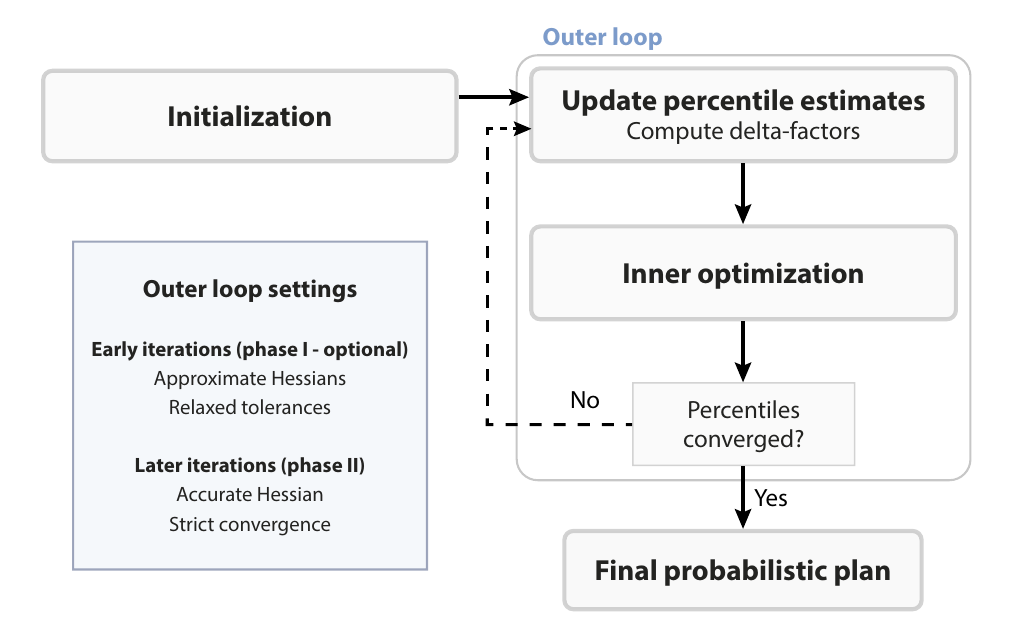}
    \caption{The probabilistic approach is sequentially solving inner optimizations. The outer loop ensures the percentile estimates remain accurate. Optionally, the outer optimization may consist of two phases, a fast search for near-optimal $\delta$-factors, followed by accurate optimization of the final beam weights.}
    \label{fig:probApproach}
\end{figure}

The structure of the probabilistic approach remains identical to previous work, shown in Figure \ref{fig:probApproach}. The optimization problem (see Section \ref{subsec:probApproach}) is solved in the inner optimization using percentile estimates with fixed $\delta$-factors. Based on the output beam weights, accurate percentiles are re-computed and checked for convergence. If not converged yet, the inner optimization is repeated with updated $\delta$-factors. The outer loop stops once the percentiles have converged.

Optionally, to reduce computation cost, one may choose to use approximate Hessians or relaxed optimality tolerances in early iterations. However, the optimization should end (last outer iterations) with accurate Hessian calculations and strict convergence to ensure that percentile convergence is reliable. The interior-point method of \verb"fmincon" \citep{matlab2024} was used for the inner optimizations, with an optimality tolerance of \num{e-5} (based on the maximum absolute Lagrangian gradient). All beam weights are initialized to 0.01.

\subsubsection{Mathematical formulation of the probabilistic optimization} \label{subsec:probApproach_mathFormulation}
The probabilistic approach of Figure \ref{fig:probApproach} is extended to using probabilistic constraints. For a given set of $\delta$-factors and objective weights $\pi_s$, the optimization is given by
\begin{eqnarray} \label{eq:probOpt_general}
    \min_{\bi{x}} & \Biggl[ \sum_{s \in S_o^{prob}}  \pi_s^{prob} f_{s,\mp}^{prob}(\bi{x},\alpha,\gamma) + \sum_{s \in S_o^{other}} \pi_s^{other} f_s^{other}(\bi{x},\gamma) \Biggr] \label{eq:objective} \\
    & \textrm{s.t.} \quad c_{i,\mp}^{prob}(\bi{x},\alpha,\gamma) \geq 0, \quad \forall s \in S_c^{prob}, \, \forall i \in I_{s} \\
    & \quad \quad \,\, c_{i}^{other}(\bi{x},\gamma) \geq 0, \quad \forall s \in S_c^{other}, \, \forall i \in I_{s}, \\
    & \quad \quad \,\, \bi{x} \geq 0,
\end{eqnarray}
where $s$ indexes structures and $i \in I_s$ denotes the active voxel set of structure $s$. The probabilistic objectives ($f^{prob}$) and constraints ($c^{prob}$) used in this work are
\begin{eqnarray}
    f_{s,\mp}^{prob}(\bi{x},\alpha,\gamma) & = \sum_{i \in I_s} \Bigl[ d_{i}^{\alpha \%}(\bi{x}) - \gamma_{i} \Bigr]_{\mp}^2, \label{eq:underoverdosage} \\
    c_{i,\mp}^{prob}(\bi{x},\alpha,\gamma) & = \mp ( \gamma_{i} - d_{i}^{\alpha \%}(\bi{x})),
\end{eqnarray}
where $[h]_{-} = \min\{ 0, h\}$ and $[h]_{+} = \max\{ 0, h\}$, so that the lower and upper signs correspond to probabilistic underdosage ($-$) and overdosage ($+$) optimization. Objectives ($f^{other}$) and constraints ($c^{other}$) labeled \textit{other} can be nominal, robust or any further type, allowing mixing of probabilistic and traditional objectives and constraints. \newline

\subsubsection{A memory-efficient percentile estimate}
\label{paragraph:memEfficient_percEstimate}

The expected dose in Equation \ref{eq:percentileEstimate} is taken from the zeroth order PCE coefficients as $\mathbb{E}[d_i(\bi{x},\bxi)] = \sum_{j \in \mathbb{B}} R_{ij}^{(0)} x_j$. The standard deviation of the voxel dose is
\begin{equation}
    \text{SD}[d_i(\bi{x},\bxi)] = \sqrt{\bi{x}^\top \bi{C}_i \bi{x}},
\end{equation}
where $\bi{C}_{i} \in \mathbb{R}^{N_b \times N_b}$ is the covariance matrix for voxel $i$. The element defined by pencil beams $j$ and $j'$ can be written in terms of PCE coefficients as
\begin{equation} \label{eq:covMatrix}
    (\bi{C}_{i})_{jj'} = \mathbb{E}[D_{ij}(\bxi)D_{ij'}(\bxi)] - \mathbb{E}[D_{ij}(\bxi)]\mathbb{E}[D_{ij'}(\bxi)] = \sum_{k=1}^P h_k^2 \, R_{ij}^{(k)} R_{ij'}^{(k)},
\end{equation}
where the $k^{th}$ PCE basis norm is $h_k^2$. As $\bi{C}_{i}$ is memory heavy ($N_b^2$ elements are computed for all voxels $i \in I_{s}$ for $s \in S_o^{prob} \cup S_c^{prob}$), we propose to use a memory-efficient reduced representation of $\bi{C}_{i}$ during optimization, by approximating $\bi{C}_{i}$ by its diagonal. As such, we obtain an approximation of $\text{SD}[d_i(\bi{x},\bxi)]$ as
\begin{equation}
    \text{SD}[d_i(\bi{x},\bxi)] = \sqrt{\bi{x}^\top \bi{C}_i \bi{x}} \approx \sqrt{ \bi{x}^\top \text{diag}(\bi{C}_i) \bi{x} } = \tilde{\sigma}_i[d_i(\bi{x},\bxi)].
\end{equation}
Here, $\text{diag}(\bi{C}_{i}) \in \mathbb{R}^{N_b}$ contains the variances of dose-influence elements $D_{ij}(\bxi)$ for voxel $i$ and beamlets $j \in 1, \ldots, N_b$. To put it differently, the $j^{th}$ element of $\text{diag}(\bi{C}_{i})$ represents the variance of dose-influence element $D_{ij}(\bxi)$ and is given by
\begin{equation} \label{eq:diagCi_j}
    \left[ \text{diag}(\bi{C}_{i}) \right]_j = (\bi{C}_{i})_{jj} = \sum_{k=1}^P h_k^2 \,\bigl( R_{ij}^{(k)} \bigr)^2,
\end{equation}
yielding
\begin{equation}
    \tilde{\sigma}_i = \sqrt{ \bi{x}^\top \text{diag}(\bi{C}_i) \bi{x} } = \sqrt{\sum_j (\bi{C}_{i})_{jj} x_j^2}.
\end{equation}
The approximate standard deviation $\tilde{\sigma}_i[d_i(\bi{x},\bxi)]$ may deviate significantly from $\text{SD}[d_i(\bi{x},\bxi)]$. However, this does not compromise the percentile estimate, because $\tilde{\sigma}_i[d_i(\bi{x},\bxi)]$ is not intended to be an SD-estimate on its own, but only meant to capture the PDF width in combination with $\delta_i$ (i.e., with estimator $\delta_i \cdot \tilde{\sigma}_i$). The $\alpha$-th percentile is approximated by $d_i^{\alpha \%}(\bi{x}) = \mathbb{E}[d_i(\bi{x},\bxi)] \pm \delta_i \cdot \tilde{\sigma}_i[d_i(\bi{x},\bxi)]$, and since the $\delta$-factors are obtained from accurate percentiles as 
\begin{equation}
    \delta_i = \pm \left( \frac{d_i^{\alpha \%}(\bi{x}) - \mathbb{E}[d_i(\bi{x},\bxi)]}{\tilde{\sigma}_i[d_i(\bi{x},\bxi)]} \right),
\end{equation}
the deviation between $\tilde{\sigma}_i$ and $\text{SD}[d_i(\bi{x},\bxi)]$ is compensated for by $\delta_i$ by construction. This compensation is maintained throughout the full optimization, as the $\delta$-factors are updated until $d_i^{\alpha \%}(\bi{x})$ has converged.

\begin{table}[b]
\caption{\label{tab:probEvaluationMetrics}Probabilistic goals (i.e., evaluation metrics) used for the CTV and dose-limiting OARs for neuro-oncological patients.}
\begin{center}
\begin{tabular}{@{}lll}
\toprule
Structure & Probabilistic goal & Equivalent percentile criterion \\[0.5ex]
\toprule
CTV               & $P(D_{99.8\%} \geq 95\% d_p) \geq 90\%$                            & 10th $D_{99.8\%} \geq 95\% d_p$                     \cr
Brainstem Core    & $P(D_{0.03\mathrm{cc}} \geq \qty{54}{\gray}\text{RBE}) \leq 5\%$   & 95th $D_{0.03\mathrm{cc}} \leq \qty{54}{\gray}\text{RBE}$ \cr
Brainstem Surface & $P(D_{0.03\mathrm{cc}} \geq \qty{60}{\gray}\text{RBE}) \leq 5\%$   & 95th $D_{0.03\mathrm{cc}} \leq \qty{60}{\gray}\text{RBE}$ \cr
Optic Nerve (L/R) & $P(D_{0.03\mathrm{cc}} \geq \qty{55}{\gray}\text{RBE}) \leq 5\%$   & 95th $D_{0.03\mathrm{cc}} \leq \qty{55}{\gray}\text{RBE}$ \cr
Optic Chiasm      & $P(D_{0.03\mathrm{cc}} \geq \qty{55}{\gray}\text{RBE}) \leq 5\%$   & 95th $D_{0.03\mathrm{cc}} \leq \qty{55}{\gray}\text{RBE}$ \cr
Retina (L/R)      & $P(D_{0.03\mathrm{cc}} \geq \qty{55}{\gray}\text{RBE}) \leq 5\%$   & 95th $D_{0.03\mathrm{cc}} \leq \qty{55}{\gray}\text{RBE}$ \cr
\bottomrule
\end{tabular}
\end{center}
\end{table}

\subsection{Choosing the probabilistic objective weights and thresholds} \label{subsec:probWeights}
For every structure that was optimized robustly (by either an objective or constraint), we defined a probabilistic equivalent (see Table \ref{tab:probEvaluationMetrics}). Target underdosage and overdosage goals are optimized to be achieved with at least 90\% probability, the probability in line with the Van Herk margin recipe \citep{vanHerk2000}. The OAR constraints defined in the EPTN guidelines \citep{Lambrecht2018} were required to be met with at least 95\% probability. As no consensus currently exists on probabilistic acceptance criteria \citep{Sterpin2024}, a 95\% acceptance probability was chosen for this study.

Although Erasmus-iCycle plans are optimized using the $\epsilon$-constraint method, weighted-sum objective weights can be obtained from the Lagrange multipliers of the Erasmus-iCycle plan \citep{Breedveld2009}. These weights are used to construct the weighted-sum approach used in the probabilistic optimization. Although the probabilistic and robust objectives are different, this provides a consistent way of choosing them to achieve near equivalent plan trade-offs. The wish-lists and their probabilistic equivalents are shown in Appendix \ref{app:wishlists}.

After plan optimization, the dose distributions of the robust and probabilistic plans are either scaled on identical probabilistic target coverage (group A) or on the most restricting probabilistic OAR constraint (group B), by probabilistic metrics as in Table \ref{tab:probEvaluationMetrics}. 
As the most limiting OAR in the robust and probabilistic plan can be different for the same patient (as the trade-off is different), the scaling could be based on different OARs. Following \citet{deJong2025}, we define probabilistic target coverage as the 10th percentile of the $D_{99.8\%}$, which we denote as 10th $D_{99.8\%}$. 

\section{Results} \label{sec:Results}

\subsection{Group A: target coverage was reached}
\begin{figure}
    \centering

    % Row 1
    \begin{subfigure}{0.7\textwidth}
        \centering
        \includegraphics[width=\linewidth]{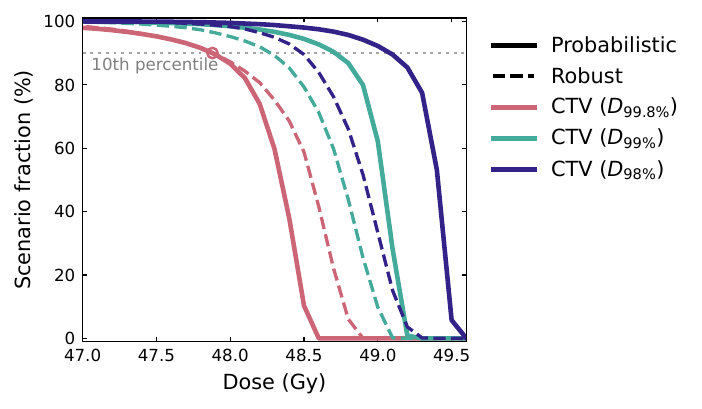}
        \caption{}
        \label{fig:DPHs_groupA_DPH}
    \end{subfigure}

    % Row 2
    \begin{subfigure}{\textwidth}
        \centering
        \includegraphics[width=0.98\linewidth]{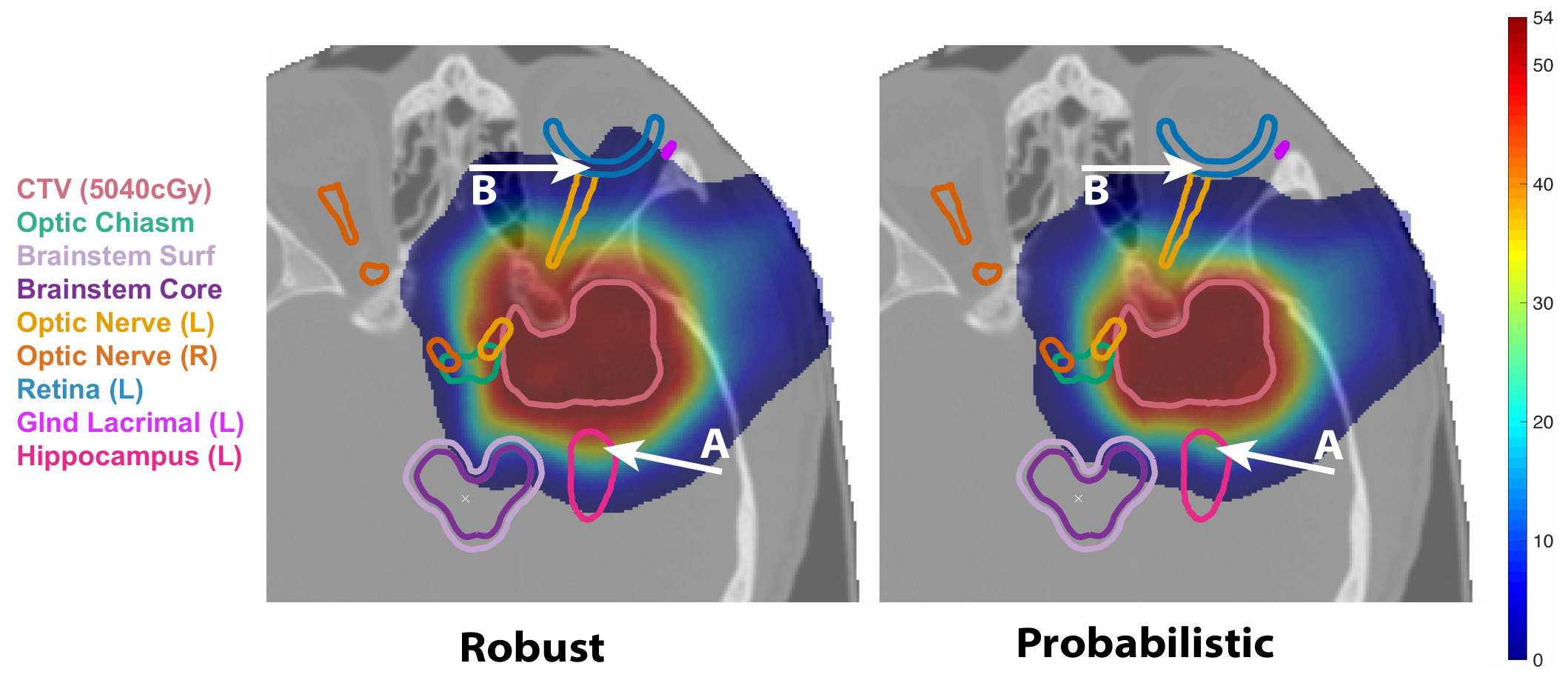}
        \caption{}
        \label{fig:DPHs_groupA_DPH_nominal}
    \end{subfigure}

    \caption{A comparison of the dose population histogram between the probabilistic and robust plan for patient 1, for probabilistic DVH metrics of the target (a), with a plan comparison in the nominal scenario (b). The probabilistic goal for target coverage (10th $D_{99.8\%} \geq 95\%d_p$) is highlighted by the pink circle. Dose values are in \unit{\gray}RBE.}
    \label{fig:DPHs_groupA}
\end{figure}

Patient 1 was scaled by identical target coverage (10th $D_{99.8\%}$), as shown the dose population histogram (DPH) in Figure \ref{fig:DPHs_groupA_DPH}. The $D_{99.8\%}$, $D_{99\%}$ and $D_{98\%}$ improved in the probabilistic plan for all scenario fractions. In the probabilistic plan, the summed dose of OAR-related DVH-metrics ($D_{0.03\mathrm{cc}}$ and $D_{40\%}$, see Table \ref{tab:OAR_reduction}) reduced by about $\qty{19}{\gray}\text{RBE}$. For clarity, we denote the difference between probabilistic and robust plans as $\Delta D_{\text{OAR,DVH}}$, giving $\Delta D_{\text{OAR,DVH}} = \qty{-19}{\gray}\text{RBE}$. The $D_{0.03\mathrm{cc},\text{OpticNerve(L)}}$ and $D_{0.03\mathrm{cc},\text{Br.Core}}$ showed a respective dose increase of \qty{0.85}{\gray}RBE and \qty{0.21}{\gray}RBE, without violating their clinical goals. The nominal mean dose decreased for all structures in Table \ref{tab:OAR_reduction_nominal}, with a total reduction of $\Delta D_{\text{OAR,nominal}} = \qty{-9.3}{\gray}\text{RBE}$.

The nominal dose distribution for patient 1 (Figure \ref{fig:DPHs_groupA_DPH_nominal}) shows that the dose extension beyond the CTV has significantly reduced compared to the robust plan. Especially reductions in the left hippocampus (arrow A: $D_{40\%}$ reduced by \qty{7.66}{\gray}) and the left retina (arrow B: $D_{0.03\mathrm{cc}}$ reduced by \qty{3.91}{\gray}) are noticeable.

\subsection{Group B: target coverage was compromised}
\begin{figure}
    \centering

    % Row 1
    \begin{subfigure}{0.49\textwidth}
        \centering
        \includegraphics[width=\linewidth]{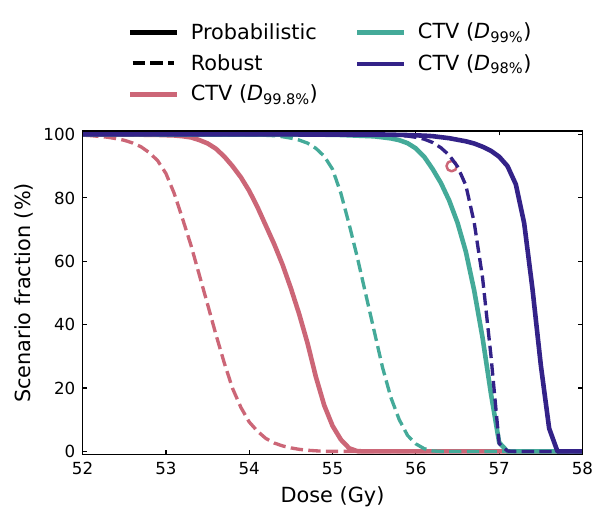}
        \caption{}
        \label{fig:DPHs_groupB_Neuro12CTV}
    \end{subfigure}
    \begin{subfigure}{0.49\textwidth}
        \centering
        \includegraphics[width=\linewidth]{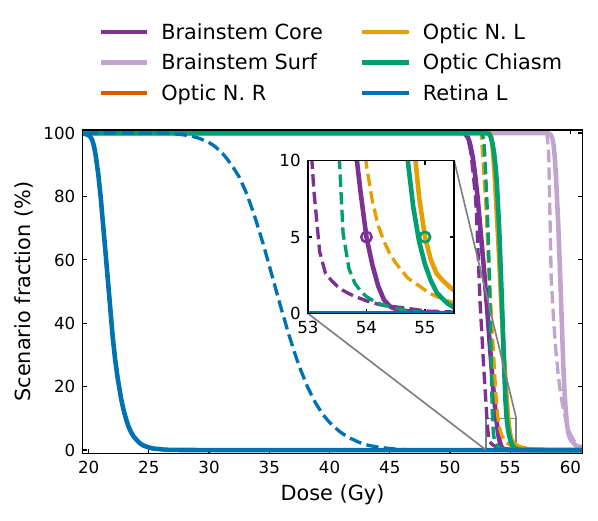}
        \caption{}
        \label{fig:DPHs_groupB_Neuro12OAR}
    \end{subfigure}

    % Row 2
    \begin{subfigure}{0.49\textwidth}
        \centering
        \includegraphics[width=\linewidth]{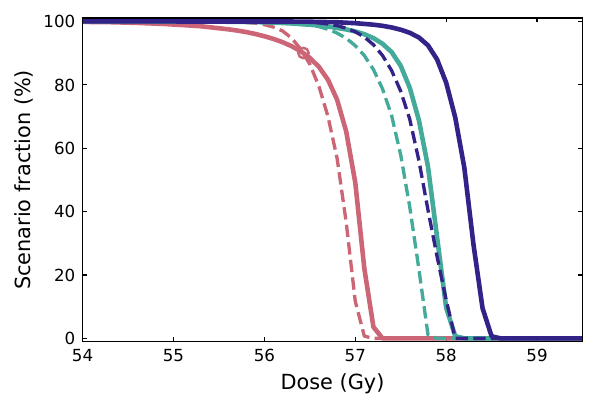}
        \caption{}
        \label{fig:DPHs_groupB_Neuro32CTV}
    \end{subfigure}
    \begin{subfigure}{0.49\textwidth}
        \centering
        \includegraphics[width=\linewidth]{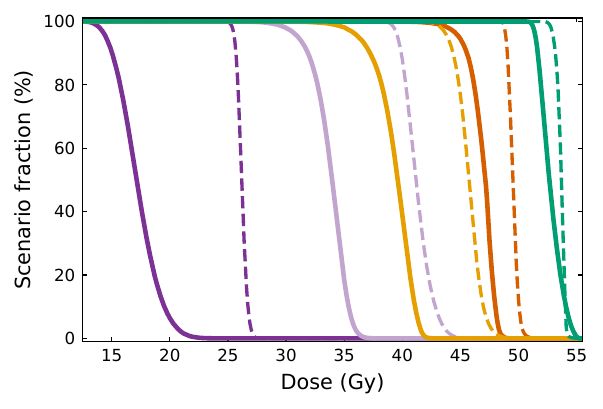}
        \caption{}
        \label{fig:DPHs_groupB_Neuro32OAR}
    \end{subfigure}
    % Neuro32_v15_target_DPH.pdf
    % Neuro32_v15_OARs_DPH.pdf
    
    % Row 3
    \begin{subfigure}{0.49\textwidth}
        \centering
        \includegraphics[width=\linewidth]{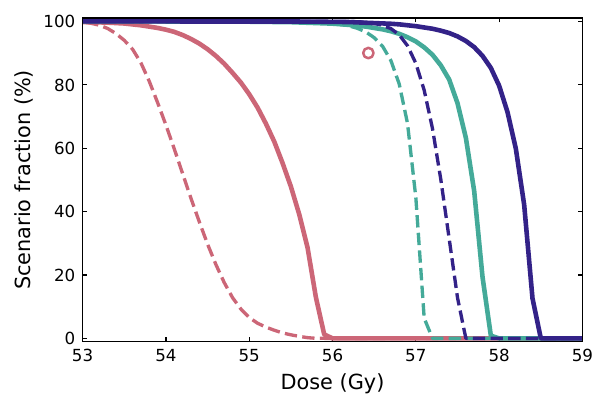}
        \caption{}
        \label{fig:DPHs_groupB_Neuro35CTV}
    \end{subfigure}
    \begin{subfigure}{0.49\textwidth}
        \centering
        \includegraphics[width=\linewidth]{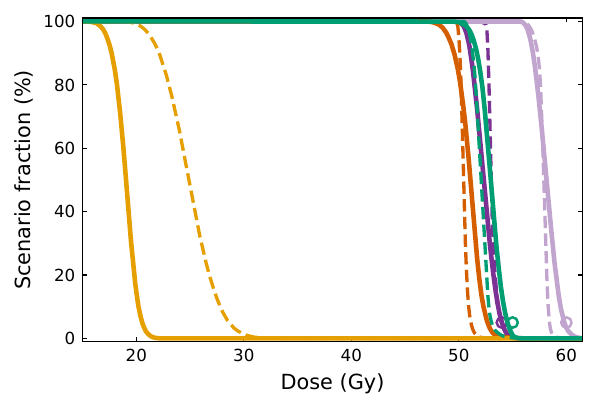}
        \caption{}
        \label{fig:DPHs_groupB_Neuro35OAR}
    \end{subfigure}

    \caption{A comparison of the dose population histograms between the probabilistic and robust plans for patients 3 (a, b), 4 (c, d) and 5 (e, f), for probabilistic DVH metrics of target ($D_{99.8\%}, D_{99\%}, D_{98\%}$, left) and OARs ($D_{0.03\mathrm{cc}}$, right). Relevant probabilistic goals are shown as circles.}
    \label{fig:DPHs_groupB}
\end{figure}

All patients in this group had insufficient \textit{clinical} target coverage, because it was limited by OARs. While the robust plan for patient 4 did not reach \textit{clinical} target coverage (as $\text{VWmin}_{CTV} \geq 95\%d_p$ only for 96.4\% of its volume - instead of 98\%), it could reach sufficient \textit{probabilistic} target coverage (see Figure \ref{fig:DPHs_groupB}). Therefore, patient 4 was scaled to have identical probabilistic target coverage, whereas the other patients were scaled to have their most limiting OAR satisfy its probabilistic constraint (Table \ref{tab:probEvaluationMetrics}).

\begin{figure}
    \centering

    % Row 1
    \begin{subfigure}{0.46\textwidth}
        \centering
        \includegraphics[width=\linewidth]{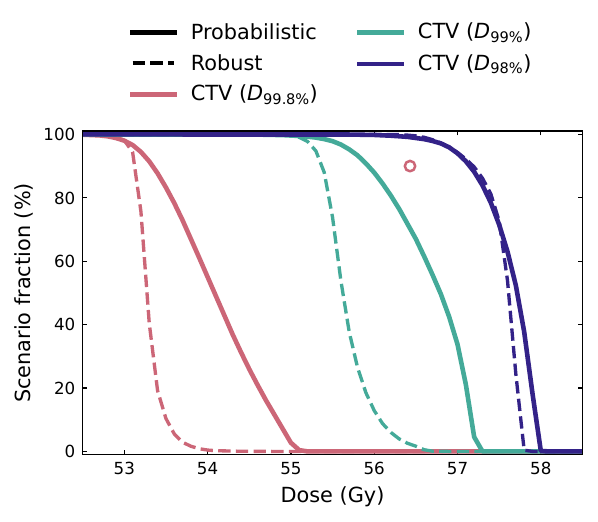}
        \caption{}
        \label{fig:groupB_Neuro10_DPH-CTV}
    \end{subfigure}
    \begin{subfigure}{0.46\textwidth}
        \centering
        \includegraphics[width=\linewidth]{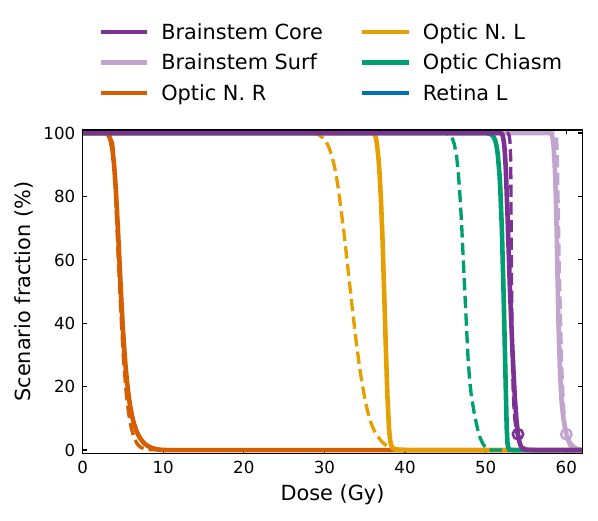}
        \caption{}
        \label{fig:groupB_Neuro10_OARdph}
    \end{subfigure}

    % Row 2
    \begin{subfigure}{0.95\textwidth}
        \centering
        \includegraphics[width=\linewidth]{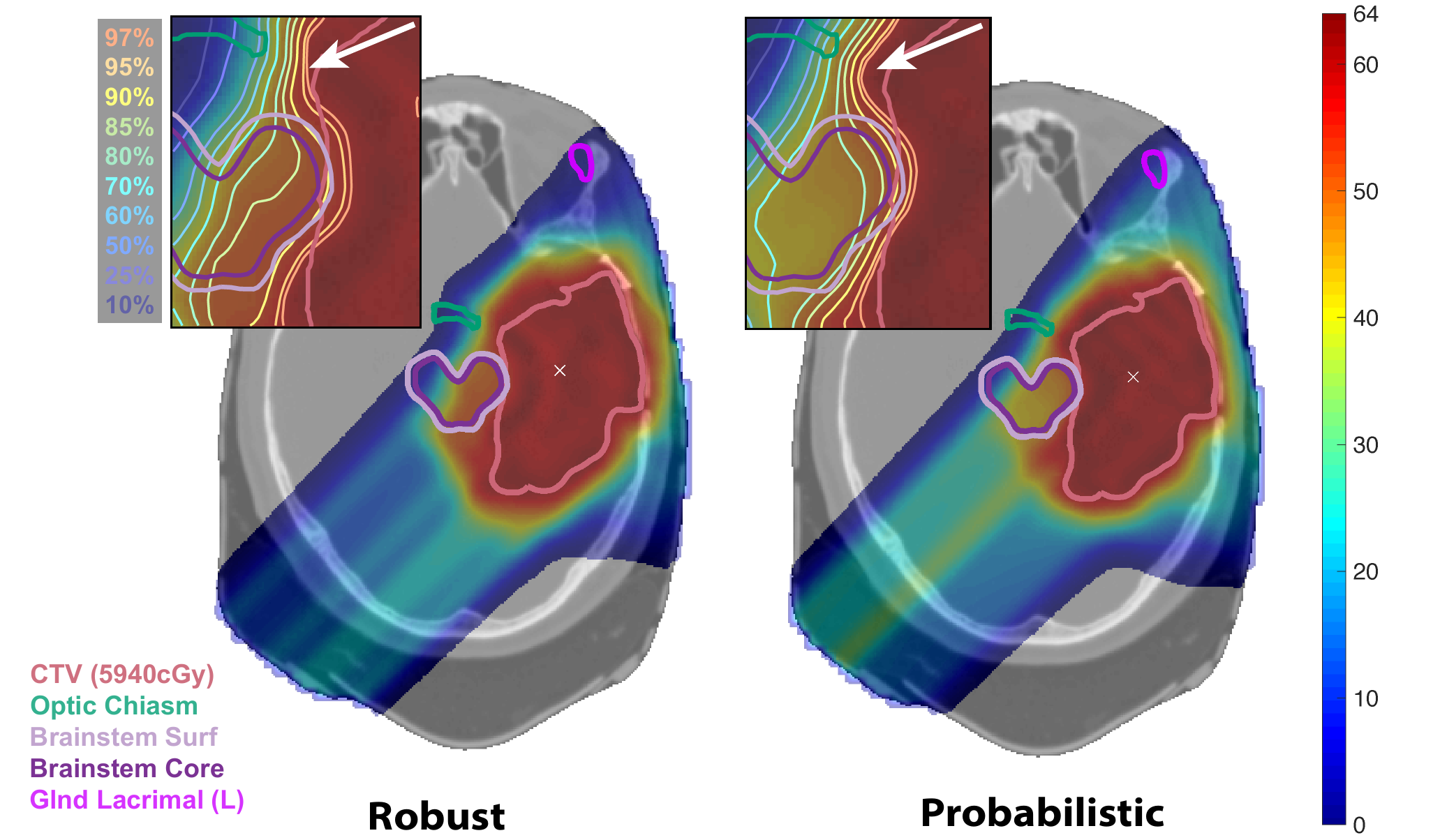}
        \caption{}
    \end{subfigure}

    \caption{A comparison of the dose population histograms between robust and probabilistic plans, for probabilistic DVH metrics of target ($D_{99.8\%}, D_{99\%}, D_{98\%}$, a) and OARs ($D_{0.03\mathrm{cc}}$, b), with a plan comparison in the nominal scenario (c) for patient 2. Relevant probabilistic goals are shown as circles and dose values are in \unit{\gray}RBE.}
    \label{fig:groupB_Neuro10}
\end{figure}

The mean increase of CTV-related DVH-metrics (10th $D_{99.8\%}$\footnote{For the 10th $D_{99.8\%}$, patient 4 was excluded because the robust and probabilistic plan were scaled to identical target coverage.}, 10th $D_{99\%}$, 10th $D_{98\%}$) across patients in the probabilistic plan is 0.66 (range: 0.19 to 0.93) \unit{\gray}RBE, 0.65 (range: 0.34 to 1.20) \unit{\gray}RBE and 0.49 (range: \num{-0.05} to 0.83) \unit{\gray}RBE, respectively. The DPH in Figure \ref{fig:groupB_Neuro10} (for patient 2 as an example) shows that the improvement in target coverage is more pronounced for different probabilities of acceptance, and the nominal dose comparison shows more conformal dose in the probabilistic plan. The summed dose of OAR-related DVH-metrics increased by $\Delta D_{\text{OAR,DVH}} = \qty{8.1}{\gray}\text{RBE}$ (Table \ref{tab:OAR_reduction}), without violating the clinical constraints. There was especially room to increase $D_{0.03\mathrm{cc},\text{OpticNerve(L)}}$ and $D_{0.03\mathrm{cc},\text{OpticChiasm}}$, to allow for improvement of CTV coverage in the neighboring CTV region (see the white arrow in Figure \ref{fig:groupB_Neuro10}). The summed nominal mean dose for patient 2 reduced by $\Delta D_{\text{OAR,nominal}} = \qty{-5.5}{\gray}\text{RBE}$ (range of individual OARs: \qtyrange{-6.8}{2.5}{\gray}RBE). Scaling the robust plan to the same probabilistic target coverage was not possible, as it was limited by the probabilistic $D_{0.03\mathrm{cc},\text{Br.Surface}}$ constraint (see Figure \ref{fig:groupB_Neuro10_OARdph}). Therefore, the improved target coverage in the probabilistic plan could not be achieved by a simple dose scaling of the robust plan.

The summed dose of OAR-related DVH-metrics for patients 3, 4 and 5 decreased, corresponding to $\Delta D_{\text{OAR,DVH}} = \qty{-33}{\gray}\text{RBE}$, $\Delta D_{\text{OAR,DVH}} = \qty{-32.6}{\gray}\text{RBE}$ and $\Delta D_{\text{OAR,DVH}} = \qty{-15.3}{\gray}\text{RBE}$, respectively. The DPHs in Figure \ref{fig:DPHs_groupB} compare the dose differences for all probabilities of acceptance (or scenario fraction). For patient 3, target coverage could be improved (Figure \ref{fig:DPHs_groupB_Neuro12CTV}) by increasing $D_{0.03\mathrm{cc},\text{OpticNerve(L)}}$, $D_{0.03\mathrm{cc},\text{OpticChiasm}}$ and $D_{0.03\mathrm{cc},\text{Br.Core}}$ to their clinical constraints (Figure \ref{fig:DPHs_groupB_Neuro12OAR}). Although the 95th $D_{0.03\mathrm{cc},\text{Br.Core}}$ increased by \qty{0.84}{\gray}RBE, the steep dose-fall off yields a decrease of \qty{8.34}{\gray}RBE for the 95th $D_{2\%,\text{Br.Core}}$ (see Appendix \ref{app:moreResults} for the DPHs). In the same plan, the sum of the 95th $D_{0.03\mathrm{cc}\text{Retina(L)}}$ and $D_{0.03\mathrm{cc}\text{Retina(R)}}$ could be reduced by about \qty{28}{\gray}RBE without compromising target coverage. For patient 5, target coverage was improved (Figure \ref{fig:DPHs_groupB_Neuro35CTV}), especially by increasing $D_{0.03\mathrm{cc},\text{Br.Surf}}$ to its clinical constraint (Figure \ref{fig:DPHs_groupB_Neuro35OAR}).

Nominal dose comparisons for patients 3, 4 and 5 are shown in Appendix \ref{app:moreResults}.

\begin{table}
\caption{\label{tab:OAR_reduction}Dose differences $\Delta D = D_{\mathrm{prob}} - D_{\mathrm{robust}}$ (\unit{\gray}RBE) of DVH metrics for 10th percentiles for CTV and 95th percentiles for OARs. Significant dose reduction ($\Delta D \leq \qty{-1.5}{\gray}\text{RBE}$) and dose increase ($\Delta D \geq \qty{1.5}{\gray}\text{RBE}$) are denoted in \textbf{bold}. Group A: target coverage met in the robust plan. Group B: OARs limited target coverage in the robust plan.}
\begin{center}
\sisetup{table-format=-2.2, detect-weight=true, detect-inline-weight=math}
\begin{tabular}{@{}ll rrrrr}
\toprule
 & & {Group A} & \multicolumn{4}{c}{Group B} \\
\cmidrule(lr){3-3} \cmidrule(lr){4-7}
Structure & Metric & {Pt\,1} & {Pt\,2} & {Pt\,3} & {Pt\,4} & {Pt\,5} \\[0.5ex]
\toprule
CTV & $D_{98\%}$   &  0.60 & -0.05 &  0.59 &  0.58 & 0.83 \\
CTV & $D_{99\%}$   &  0.44 &  0.54 &  1.20 &  0.34 & 0.52 \\
CTV & $D_{99.8\%}$ &  0.00 &  0.19 &  0.86 &  0.00 & 0.93 \\
\midrule
Brainstem Core    & $D_{0.03\mathrm{cc}}$ &           0.21  &          0.12  &            0.84  & \bfseries -6.63  &           0.00  \\
Brainstem Surface & $D_{0.03\mathrm{cc}}$ & \bfseries -3.05 &          0.00  &           -0.15  & \bfseries -7.62  &           1.37  \\
Optic Chiasm      & $D_{0.03\mathrm{cc}}$ &          -1.20  & \bfseries 3.43 &            1.29  &           0.35   &           1.47  \\
Optic Nerve (L)   & $D_{0.03\mathrm{cc}}$ &           0.85  & \bfseries 1.65 &            0.74  & \bfseries -5.98  & \bfseries -8.09 \\
Optic Nerve (R)   & $D_{0.03\mathrm{cc}}$ & \bfseries -2.76 &          0.58  &            0.58  & \bfseries -2.01  & \bfseries  1.54 \\
Lens (L)          & $D_{0.03\mathrm{cc}}$ &          -0.15  &         -0.03  & \bfseries -2.49  &           0.10   &          -0.07  \\
Lens (R)          & $D_{0.03\mathrm{cc}}$ &           0.00  &          0.00  &           -0.99  &           0.44   &           1.31  \\
Retina (L)        & $D_{0.03\mathrm{cc}}$ & \bfseries -3.91 &          0.01  & \bfseries -17.33 &           0.26   &          -0.68  \\
Retina (R)        & $D_{0.03\mathrm{cc}}$ &           0.01  &    {--}       & \bfseries -10.63 & \bfseries -6.76  & \bfseries -4.47 \\
Cornea (L)        & $D_{0.03\mathrm{cc}}$ &          -1.36  &     {--}      & \bfseries -3.09  &           0.09   &          -0.18  \\
Cornea (R)        & $D_{0.03\mathrm{cc}}$ &           0.01  &    {--}       &           -1.27  &           0.41   &           1.21  \\
Hippocampus (L)   & $D_{40\%}$            & \bfseries -7.66 &          0.37  &           -0.55  &          -0.95   &          -0.04  \\
Hippocampus (R)   & $D_{40\%}$            &           0.00  & \bfseries 1.92 &            0.02  & \bfseries -4.30  & \bfseries -8.70 \\
\midrule
Total $\Delta D_{\text{OAR,DVH}}$ & & \bfseries -19.00 & \bfseries 8.05 & \bfseries -33.03 & \bfseries -32.61 & \bfseries -15.32 \\
\bottomrule
\end{tabular}
\end{center}
\end{table}

\begin{table}
\caption{\label{tab:OAR_reduction_nominal}Dose differences $\Delta D = D_{\mathrm{prob}} - D_{\mathrm{robust}}$ (\unit{\gray}RBE) for the nominal scenario of OARs evaluated by mean dose. Significant dose reduction ($\Delta D \leq \qty{-1.5}{\gray}\text{RBE}$) and dose increase ($\Delta D \geq \qty{1.5}{\gray}\text{RBE}$) are denoted in \textbf{bold}. Group A: target coverage met in the robust plan. Group B: OARs limited target coverage in the robust plan.}
\begin{center}
\sisetup{table-format=-2.2, detect-weight=true, detect-inline-weight=math}
\begin{tabular}{@{}ll rrrrr}
\toprule
 & & {Group A} & \multicolumn{4}{c}{Group B} \\
\cmidrule(lr){3-3} \cmidrule(lr){4-7}
Structure & Metric & {Pt\,1} & {Pt\,2} & {Pt\,3} & {Pt\,4} & {Pt\,5} \\[0.5ex]
\toprule
Brain-CTV          & Mean &          -0.97  & \bfseries -1.64 &          -1.19  &          -1.26  & \bfseries -1.60 \\
Cerebrum-CTV       & Mean &          -1.09  &          -1.48  &          -1.27  &          -1.42  & \bfseries -1.83 \\
Cochlea (L)        & Mean &          -0.02  & \bfseries -6.75 & \bfseries -3.51 &           0.00  &           0.00  \\
Cochlea (R)        & Mean &           0.00  &          -0.01  &           0.00  &          -0.05  &          -0.28  \\
Lacrimal gland (L) & Mean & \bfseries -2.30 & \bfseries  2.51 & \bfseries -5.69 &           0.02  &          -0.03  \\
Lacrimal gland (R) & Mean &           0.00  &           0.00  &          -0.30  &           0.13  & \bfseries  5.99 \\
Hippocampus-CTV    & Mean & \bfseries -2.47 & \bfseries  2.37 & \bfseries  1.71 &          -1.25  & \bfseries -1.71 \\
Pituitary          & Mean & \bfseries -2.48 & \bfseries -2.16 &           0.46  & \bfseries -7.72 &           0.17  \\
\midrule
Total $\Delta D_{\text{OAR,nominal}}$ & & \bfseries -9.33 & \bfseries -7.16 & \bfseries -9.79 & \bfseries -11.56 & 0.72 \\
\bottomrule
\end{tabular}
\end{center}
\end{table}

\subsection{Dose calculation and optimization times} \label{sec:timings}

Dose deposition matrices were calculated in five batches, each using two cores on Intel Xeon E5-2690 (2.90 GHz) CPUs. Taking patient 2 as a representative case (with 30174 active voxels for probabilistic objectives and constraints), the computation time per scenario was about \qty{410}{\s} (or about \qty{40}{\milli\s} per beamlet). $D_{ij}$ PCEs of all relevant structures were constructed sequentially and took \qty{5134}{\s} (excluding dose calculation), i.e., about \qty{170}{\s} per 1000 voxels.

Table \ref{tab:optTimes} summarizes total optimization time with the computation times of sampling, percentile calculation and PCE conversion (see Equation \ref{eq:Methods_PCEdose} in Appendix \ref{app:PCEaccuracy}). The mean total probabilistic optimization time was \qty{107}{\hour}. 3/5 probabilistic optimizations were done within \qty{80}{\hour}, with two exceptions taking \qty{141}{\hour} and \qty{226}{\hour}. Inner optimization time is dominating total optimization time (contributing more than 98\% for all patients), whereas sampling time ($\leq0.1\%$) and percentile calculation ($\leq1.1\%$) contributed only slightly. Optimization times could be reduced retrospectively from 16\% to 72\% by choosing different convergence criteria (see Section \ref{sec:Discussion}).

\begin{table}
\caption{\label{tab:optTimes}Total and inner optimization times (in hours) of the probabilistic plans, with the computation times of sampling, percentile calculation and PCE conversion (in seconds).}
\begin{center}
\begin{tabular}{@{}lccccc}
\toprule
 & Pt.\ 1 & Pt.\ 2 & Pt.\ 3 & Pt.\ 4 & Pt.\ 5 \\[0.5ex]
\toprule
Inner optimization (\unit{\hour})  & 43.6 & 223  & 45.8 & 79.3 & 138  \\
Sampling (\unit{\second})                & 69   & 645  & 237  & 114  & 636  \\
Percentile calculation (\unit{\second})  & 356  & 5556 & 634  & 645  & 3145 \\
PCE conversion (\unit{\second})          & 307  & 3042 & 785  & 510  & 5655 \\
\midrule
\textbf{Total optimization} (\unit{\hour}) & \textbf{43.9} & \textbf{226} & \textbf{46.5} & \textbf{79.8} & \textbf{141} \\
\bottomrule
\end{tabular}
\end{center}
\end{table}
\section{Discussion} \label{sec:Discussion}

This study shows the feasibility of our previously proposed probabilistic approach \citep{JRdeJong2026} on five neuro-oncological patients. 

For group A, probabilistic planning reduced the summed dose of OAR-related DVH-metrics, corresponding to $\Delta D_{\text{OAR,DVH}} = \qty{-19}{\gray}\text{RBE}$, while reaching identical probabilistic target coverage. The reason is that the robust voxel-wise CTV underdosage constraint requires all voxels to satisfy the dose threshold in all 21 scenarios, while for the probabilistic approach coverage has to be achieved in 90\% of the cases. Scenarios with significant displacements (e.g., $2.5 \sigma_{setup} = \qty{3}{\milli\meter}$) from the nominal scenario are unlikely to occur, but in robust planning are considered equally important as the nominal scenario. 

Patients in group B that could not achieve probabilistic robustness in their robustly optimized plans (patient 2, 3 and 5) have improved target coverage in the probabilistic plan, at the expense of dose increase in OARs (but within clinical constraints). Patient 4 reached sufficient \textit{probabilistic} target coverage in both plans, but probabilistic planning improved the DVH-metrics of CTV and OAR further. This is consistent with the DPH of $D_{99.8\%}$ for the probabilistic plan (Figure \ref{fig:DPHs_groupB_Neuro32CTV}), which extends to lower dose values above 90\% scenario fraction. The small fraction of scenarios ($<10\%$) that leads to lower target dose may contribute to the additional OAR sparing.

In patients 3, 4 and 5, the probabilistic approach allowed to reduce CTV underdosage by optimizing the probabilistic OAR constraints to their goals (see Figure \ref{fig:DPHs_groupB_Neuro12OAR}), at the same time reducing OAR dose if possible (e.g., if they are not dose limiting, like the left and right retina for patient 3). \newline

The main purpose of this study was to show feasibility on actual patient cases, but some clinical details differ from current clinical practice for these patients. The clinical RayStation plans were used as a starting point for generating the robust and probabilistic plans, but the benchmark robust plans were not identical to the clinical RayStation plans. Robust and probabilistic planning was done using MFO, whereas single-field optimization (SFO) is currently used for neurological tumors at HollandPTC. Beam settings were taken from the clinical SFO plans, meaning that beam directions were not optimized for MFO, which has potentially led to underestimated plan quality. To understand the full potential of probabilistic optimization with MFO, we may extend to using optimized beam angle configurations in the future.

Systematic errors in the probabilistic optimization were chosen to be consistent with the Van Herk margin recipe. Their standard deviations were slightly overestimated, because random errors were not considered. In fact, SDs of setup errors at HollandPTC were found to be \qty{0.7}{\milli\meter} instead of \qty{1.2}{\milli\meter} \citep{RojoSantiago2021}, which may lead to conservative dose extensions beyond the target. Clinical values (of 3\%) were chosen for the SD of the range error (as in \citet{deJong2025}) -- whereas lower values have also been suggested in literature \citep{Wohlfahrt2019} -- potentially leading to too conservative robust plans.  When focusing on more clinical comparisons in future work, both SDs should be carefully reviewed. \newline

Probabilistic optimization approaches have been proposed for photon planning by optimizing probabilistic minimum target dose \citep{Gordon2010,Mescher2017} for systematic and random errors. \citet{Tilly2019} developed a probabilistic approach for optimizing pDVH metrics for photon planning under systematic setup errors and geometric uncertainty, based on a CVaR description (using \num{100} scenarios) of a dose parameter that was empirically tuned for the $D_{98\%}$. More recently \citep{Fredriksson2026}, the use of the \textit{percentile-trimmed power mean} was proposed for systematic and random errors, which -- depending on the parameters -- is a trade-off between the expected value and exact percentile. Although optimization of an exact percentile may be numerically difficult using this metric, the method showed improved probabilistic trade-offs compared to clinical standards for a Volumetric Modulated Arc Therapy (prostate) and proton therapy (brain) case.

Our probabilistic approach builds on these previous percentile-based optimization studies for systematic setup errors, but differs in how uncertainties are included in the optimization. By using PCE for uncertainty quantification of the dose-influence matrix, the approach is able to capture non-linear dose dependencies (and thus their corresponding non-Gaussian PDFs). This dose dependence is translated towards corresponding percentiles by repeatedly updating the $\delta$-factors. In practice, this means that the optimization is not only steered into the desired direction (as was done before by assuming Gaussian-distributed dose \citep{Sobotta2010,Fabiano2022}), but can precisely optimize towards a personalized probabilistic goal. 

This work is a first step in comparing direct probabilistic optimization as alternative to robust optimization. \citet{deJong2025} has shown that adjusting the robustness settings for robust optimization, followed by rescaling, already improves plan quality from a probabilistic perspective. Although that study was applied to the same neuro-oncological patient group, making a fair comparison to that is difficult, as the dose engine and optimization method are different. Also random errors were not considered in the current work. Therefore, the advantages of direct probabilistic optimization versus indirect probabilistic optimization \citep{deJong2025} remain to be investigated. \newline

\mycomment{
Indirect optimization methods may provide an effective first step to transition the clinical standard to using probabilistic evaluation and optimization. Re-optimization of clinical (mini-max robust) plans based on probabilistic (IMPT) plan evaluation \citep{deJong2025} has shown to either improve target coverage, or reduce organ dose where possible. Although that study was applied to the same neuro-oncological patient group, a fair comparison cannot be done, as the dose engine and optimization method are different. Also random errors were not considered in this work. Alternatively, \textit{robustness recipes} \citep{vanderVoort2016} can be used to determine which error scenarios should be included into the clinically used robust optimization to ensure probabilistic target coverage. However, a general recipe does not account for personalized trade-offs between target and nearby located organs. \newline
}

Probabilistic planning was done using voxel-wise percentiles of dose, whereas probabilistic evaluation was based on pDVH metrics (e.g., 10th $D_{99.8\%}$ or 95th $D_{0.03\mathrm{cc}}$). This introduces a mismatch between optimization and evaluation, leading to indirect control over evaluation metrics. To match the pDVH metrics to the clinical constraints, probabilistic goals were adjusted by changing the dose threshold. Similar tuning was done by \citet{Tilly2019} to match optimization parameters to pDVH metrics. In this work, we restricted ourselves to achieve the probabilistic goals by at least 90\% (target) or 95\% (OAR), which were chosen heuristically. Different probabilities may have led to better correspondence with the pDVH metrics, thereby improving direct control over the probabilistic evaluation. In general, trade-offs can potentially be improved by tuning the objective weights, or by doing a sequential probabilistic optimization similar to prioritized optimization approaches such as Erasmus-iCycle, so that objective weights do not have to be tuned. \newline

Voxel selection of structures may affect the probabilistic outcome. For large structures, a limited number of voxels was included in the optimization (the \textit{active} set), meaning that all \textit{inactive} voxels were not penalized. This effect is not inherent to probabilistic optimization, but a direct consequence of voxel selection in any voxel-based optimization. As a result, probabilistic goals were only enforced in \textit{active} voxels, and may not be met for some \textit{inactive} voxels. 

For example, we have seen that if too few voxels are included in dose-limiting OARs (e.g., the brainstem -- constrained using the 95th percentile of its voxel dose), inactive brainstem voxels could reach doses above the constraint threshold. Similarly, inactive target voxels could become under- or overdosed. Therefore, for some patients, the thresholds corresponding to the probabilistic goals were slightly adjusted to control the dose in inactive voxels. The effect would diminish (in both the robust and probabilistic plans) if we increased the voxel sampling density. However, that would increase computational cost, especially for large structures. Because probabilistic optimization can particularly improve tradeoffs in regions where target and OAR lie close to each other, both a thin inner CTV edge and regions of CTV and OAR that are close or overlapping were always fully included in the wish-list. \newline

For robust planning, dose-influence matrices $D_{ij}^{r}$ were computed and stored for $r = 1, \ldots, N_r$ robust scenarios (in this work, $N_r = 21$). For the probabilistic approach, the $D_{ij}(\bxi)$ PCE requires most memory, as PCE coefficients $R_{ij}^{(k)} \in \mathbb{R}^{N_v \times N_b}$ must be stored for $k = 0, \ldots, P$, in this work corresponding to $P+1 = 73$ basis vectors. The expected dose-influence matrix $\mathbb{E}[D_{ij}(\bxi)] = R_{ij}^{(0)}$ as well as the reduced covariance term (see Equation \ref{eq:diagCi_j}) could directly be obtained from the PCE coefficients. Therefore, robust optimization is approximately a factor $( P+1 )/ N_r = 73/21 \approx 3.5$ more memory-efficient. However, once the PCE coefficients have been constructed, we expect the matrix operations within the inner iterations to be comparable to those in robust optimization, as only $\mathbb{E}[D_{ij}(\bxi)]$ and $\text{diag}(\bi{C}_i)$ are involved.

Dose calculation times in this study are significant and could be reduced by optimizing the dose engine in terms of speed, for example by using graphics processing units (GPUs) for acceleration \citep{Feng2024,Kalantzis2015,Ma2014}. Computation time would also be reduced by parallelizing over structures. The use of PCE however is not inherent to the probabilistic planning approach, so other efficient sampling (or dose calculation) methods can be used instead, for example by methods that are deep-learning based \citep{StryjaM,PastorSerrano2022}. Nevertheless, PCE was preferred in this approach because it allows us to obtain analytical expressions for the expected dose, dose variance and their gradients. Percentiles were obtained from the PCE in about \qty{3.6}{\second} per \num{1000} voxels on average, of which about \qty{0.5}{\second} was due to sampling the \num{100.000} error scenarios. \newline

Throughout this study, convergence settings were chosen to be conservative to ensure full percentile convergence for all patient cases. This partly contributed to the significant optimization times reported in Table \ref{tab:optTimes}. To quantify the potential time reduction that can be achieved by using less strict convergence settings, we retrospectively relaxed the convergence criteria (see Appendix \ref{app:convCriteria} for a full analysis). For patient 1, 2 and 3, the total optimization could be reduced by 26\%, 72\% and 16\% respectively, while maintaining comparable plan quality\footnote{The 95th percentile of OAR-related DVH metrics (Table \ref{tab:OAR_reduction}) stayed below 1\% (for all metrics above \qty{0.02}{\gray}RBE).}. This analysis illustrates that the optimization time could be sensitive to the choice of convergence criteria.

Inner optimization times are dominating the total optimization time for all patients, possibly due to the following factors. 
\begin{enumerate}
    \item Probabilistic terms optimize for all \textit{active} voxels, whereas further computational gain is expected if the objectives and constraints are rewritten such that only the most (or for example the 5\% most) violating voxels within the active voxel set are penalized. This implementation would be similar as done in Erasmus-iCycle, where for some type of robust objectives and constraints only the worst voxel was penalized. 
    \item The probabilistic approach relies significantly on warm-starting between successive (outer) iterations. Interior-point methods, as used in the inner optimization, are known for having potential warm-starting difficulties \citep{Aleman2010,Gondzio2008}. Warm-starting may therefore have limited computational benefits in the current approach.
\end{enumerate}
To quantify the computational overhead due to warm-starting, we performed two additional optimizations (see Appendix \ref{app:convCriteria}), where we updated the $\delta$-factors during the inner optimization (for every 10 iterations) rather than by an outer-loop (i.e., by warm-starting). As a result, initialization of the optimization was only needed once, instead of for every outer iteration. For patient 4, the total optimization time could be reduced from \qty{79.8}{\hour} to \qty{9.8}{\hour} (i.e., by 88\%). For patient 5, the total optimization time could be reduced from \qty{141}{\hour} to \qty{8.6}{\hour} (i.e., by 94\%). \newline

\begin{table}
\caption{\label{tab:optTimesReduction}Total optimization times in hours of the probabilistic plans, showing the potential time reductions by either relaxing the convergence criteria, or avoiding warm-starting (i.e., without outer loop).}
\begin{center}
\begin{tabular}{@{}lccccc}
\toprule
 & Pt.\ 1 & Pt.\ 2 & Pt.\ 3 & Pt.\ 4 & Pt.\ 5 \\[0.5ex]
\toprule
Tot.\ optimization (default) (\unit{\hour})              & 43.9 & 226  & 46.5 & 79.8 & 141 \\
\midrule
Tot.\ optimization (relaxed convergence) (\unit{\hour})  & 32.2 & 62.8 & 39.1 & --   & --  \\
Tot.\ optimization (without outer loop) (\unit{\hour})   & --   & --   & --   & 9.8  & 8.6 \\
\bottomrule
\end{tabular}
\end{center}
\end{table}

The presented probabilistic planning approach is sufficiently general to be applied to other radiotherapy modalities. The approach may be beneficial as well for therapies that have a lower sensitivity to uncertainties (e.g., photon therapy), but that still produce sharp dose gradients in their treatment plans. This is especially relevant for anatomies where similar tradeoffs between target coverage and OAR sparing appear, such as pancreatic cancer treated with stereotactic body radiotherapy \citep{Loi2019}. Treating uncertainties probabilistically, rather than by margins, could potentially reduce organ dose. 

The probabilistic planning approach could be extended towards including random setup errors, so that we could probabilistically optimize for (hypo-)fractionated treatments. Including fractionation has been studied before in photon \citep{Unkelbach2004,Unkelbach2018} and proton planning \citep{Wahl2018} for robust objectives involving expected dose and dose variance. Particularly interesting would be to understand how the plan quality gets affected by including the number of fractions explicitly in the percentile-based optimization. Further studies may also look into the feasibility of including anatomical variations and contour uncertainties into the optimization. Treating the target as a probabilistic map rather than a binary volume would allow to account for uncertainties in tumor extent \citep{Buti2021}.

Additionally, extending the probabilistic planning approach towards optimizing for pDVH metrics would allow for direct optimization of clinically relevant evaluation metrics. 

\section{Conclusions} \label{sec:Conclusions}
Clinical feasibility of the previously proposed probabilistic approach \citep{JRdeJong2026} was demonstrated on five neuro-oncological patients. Probabilistic planning has been shown to give promising anatomy-specific trade-offs between target coverage and OAR sparing: it reduced excessive dose where possible (keeping coverage with a certain probability) and simultaneously improved target robustness close to OARs. The desired amount of robustness could be controlled by defining personalized probabilistic goals on voxel-wise percentiles. Future work should include random errors, extend towards probabilistic DVH optimization and evaluate its applicability to other treatment modalities that may benefit from probabilistic optimization.
\section*{Data availability statement}

The Open Source Generalized Polynomial Chaos Expansion Toolbox \citep{Perk2014} (\url{https://gitlab.com/zperko/opengpc/-/tree/4a2e11cc77617c705f8a555996aee3c7cd4a61db/}) is used to construct and evaluate Polynomial Chaos Expansions. The data that support the findings of this study are subject to regulatory restrictions and therefore cannot be made publicly available upon publication. However, the data that support the findings of this study are available from the authors upon reasonable request.

\section*{Acknowledgements}

The authors would like to acknowledge that this work is funded by RaySearch Laboratories. Zoltán Perkó is an associate professor at TU Delft and is employed as a Senior Applied Scientist at Radformation Inc.. His industry employment is unrelated to the submitted work. The remaining authors have no conflicts of interest to declare.
\section*{CRediT author and contributor statement}

\textbf{Jelte R de Jong}: Conceptualization, Data Curation, Formal Analysis, Investigation, Methodology, Software, Validation, Visualization, Writing - Original Draft, Writing - Review and Editing \\
\textbf{Sebastiaan Breedveld}: Conceptualization, Resources, Software, Writing - Review and Editing \\
\textbf{Steven Habraken}: Conceptualization, Writing - Review and Editing \\
\textbf{Mischa Hoogeman}: Conceptualization, Funding Acquisition, Writing - Review and Editing \\
\textbf{Jenneke I de Jong}: Conceptualization, Resources, Writing - Review and Editing \\
\textbf{Danny Lathouwers}: Conceptualization, Methodology, Project Administration, Resources, Supervision, Writing - Review and Editing \\
\textbf{Zoltán Perkó}: Conceptualization, Funding Acquisition, Methodology, Project Administration, Resources, Supervision, Writing - Review and Editing

% Conceptualization, Data Curation, Formal Analysis, Funding Acquisition, Investigation, Methodology, Project Administration, Resources, Software, Supervision, Validation, Visualization, Writing - Original Draft, Writing - Review and Editing

\bibliography{myrefs}

\clearpage
\appendix
\newpage

\section{Accuracy analysis of the PCE} \label{app:PCEaccuracy}

Probabilistic plan evaluation is done based on constructing a PCE of the voxel dose of the final plan (Equation \ref{eq:diResponses}). The PCE gives an approximation of the dose engine as function of uncertainty, in this work for setup errors in X, Y and Z (mean: \qty{0.0}{\milli\meter}, SD: \qty{1.2}{\milli\meter}) and a range error (mean: 1.2\%, SD: 1\%). The nominal scenario therefore corresponds to a range error of 1.2\% and \qty{0}{\milli\meter} setup errors. In the following, we consider 2 PCEs: a) the PCE used for plan evaluation (the voxel dose PCE, or $d_i$ PCE), and b) the PCE used during optimization (the dose influence PCE, or $D_{ij}$ PCE), which is converted to a voxel dose PCE ($D_{ij} \rightarrow d_i$) by Equation \ref{eq:Methods_PCEdose}. In this accuracy analysis, we consider the voxel dose PCE (and therefore its generated percentiles) to be true values. \\

In previous studies \citep{RojoSantiago2021, Perk2016}, it was shown that an optimal trade-off between voxel dose PCE accuracy and construction cost is obtained for grid level $\text{GL} = 4$ and polynomial order $\text{PO} = 5$, where the grid level is extended (E) to $\text{GL}=5$ along the principal axes. We therefore use the same settings in this work. For this choice of grid level, 217 voxel dose inputs must be calculated. Additionally, 9 voxel dose scenarios (for the nominal scenario and the $\pm 3 \sigma$ displacements along the principal axes) are calculated to determine for which voxels a PCE will be constructed. A voxel dose PCE is constructed if its dose value exceeds threshold $\Theta_{d_i} = \qty{0.01}{\gray}\text{RBE}$ in any of the 9 scenarios. In the following we show that using identical settings for the construction of the $D_{ij} \rightarrow d_i$ PCE is sufficient for the probabilistic optimizations done in this study.

During optimization, we convert the $D_{ij}$ PCE to a voxel dose PCE (referred to as $D_{ij} \rightarrow d_i$) by
\begin{eqnarray} \label{eq:Methods_PCEdose}
    d_i(\bi{x},\bxi) & = \sum_{j \in \mathbb{B}} D_{ij}(\bxi) x_j \approx \sum_{j \in \mathbb{B}} \left( \sum_{k=0}^{P} R_{ij}^{(k)} \Psi_k(\bxi) \right) x_j \\
    & = \sum_{k=0}^{P} \left( \sum_{j \in \mathbb{B}} R_{ij}^{(k)} x_j \right) \Psi_k(\bxi) = \sum_{k=0}^{P} q_{i}^{(k)} \Psi_k(\bxi),
\end{eqnarray}
meaning the voxel dose coefficients ($q_{i}$) are obtained from the $D_{ij}$ coefficients ($R_{ij}^{(k)}$) by performing a weighted sum with the beam weights. By including too few $D_{ij}$ elements in the PCE (i.e., $R_{ij}$ would have many zeros), $q_{i}$ could become biased (because potentially non-zero PCE coefficients would be ignored). PCE coefficients can be positive and negative, so it is not trivial to predict if the voxel dose PCE would under- or overestimate the dose engine. We found that a dose threshold of $\Theta_{D_{ij}} = \qty{e-3}{\gray}\text{RBE}$ is a good tradeoff between memory cost and PCE (and percentile) accuracy. \\

The $D_{ij}$ PCE and $D_{ij} \rightarrow d_i$ PCE were constructed using the same settings as for the voxel dose PCE, except for dose threshold $\Theta$. Even if $\Theta_{D_{ij}} = \qty{0}{\gray}\text{RBE}$ would have been used (with all other settings being identical), identical PCE responses are not necessarily expected. This is because the dose distribution of a treatment plan tends to be smoother than a single $D_{ij}$ element, making the dose-influence matrix potentially more challenging to approximate. As the $D_{ij} \rightarrow d_i$ PCE is used for percentile calculation during optimization, its accuracy is checked especially regarding this aspect (see Appendix \ref{subsec:percentileAccuracy}). In Appendix \ref{subsec:PCEresponseComparison} we compare some $D_{ij} \rightarrow d_i$ PCE responses with the $d_i$ PCE responses. 

\subsection{Analysis of percentile accuracy} \label{subsec:percentileAccuracy}
In the following we analyze the percentile accuracy of the $D_{ij} \rightarrow d_i$ PCE. For this we use the probabilistic plan results of patient 4. A $D_{ij}$ PCE was constructed and converted using Equation \ref{eq:Methods_PCEdose}, so that the $D_{ij} \rightarrow d_i$ PCE was obtained. We compute percentiles by sampling the $D_{ij} \rightarrow d_i$ $100.000$ times and check these against percentiles that were sampled using the voxel dose PCE. As the $D_{ij}$ PCE was only constructed for certain \textit{active} voxels for certain structures, we only compare for these voxels and structures. Because probabilistic optimization was done using 10th and 90th percentiles for targets and with 95th percentiles for OARs, we only compare the corresponding percentiles. \\

Table \ref{tab:ptileDifferences} reports the percentile differences between the $D_{ij} \rightarrow d_i$ and voxel dose PCE for the active voxels in the corresponding structures. In particular, the mean percentile difference (over the structure voxels) with the SD of the percentile difference is shown, with the range of the percentile difference. For percentiles that are used for optimizing target underdosage (e.g., the 10th percentile), we especially aim to prevent overestimation (positive differences), as these cases make the optimizer think that the probabilistic goal is achieved whereas it is not. The largest 10th percentile overestimation for the CTV is \qty{1.16}{\gray}RBE is considered acceptable, as it is less than 2\% of its prescribed dose.

For target and organ overdosage (the 90th and 95th percentiles), we should especially aim to prevent percentile underestimation. The 90th percentile of the CTV is underestimated by \qty{0.62}{\gray}RBE (about 1\% of its prescribed dose); in the most underestimated voxel the 95th percentile is underestimated by \qty{-0.74}{\gray}RBE, which is about $1.3\%$ of its constraint threshold (\qty{55}{\gray}RBE). \newline

These biases in percentile calculation could potentially lead to contradicting voxel dose constraints. For example, assuming the largest deviations from Table \ref{tab:ptileDifferences}, a probabilistic overdosage constraint with 95\% acceptance probability for the right optic nerve at \qty{52}{\gray}RBE would be too restricting in combination with a probabilistic underdosage constraint of the CTV with 90\% acceptance at $d_p = \qty{50.4}{\gray}\text{RBE}$. However, for the patients in this work, these issues did not occur.

\begin{table}
\caption{\label{tab:ptileDifferences}Mean and standard deviation (and their range) of percentile differences between the $D_{ij} \rightarrow d_i$ and voxel dose PCE for the active voxels in the structure.}
\begin{center}
\begin{tabular}{@{}ll S[table-format=-1.3] S[table-format=1.2] c}
\toprule
Structure & Percentile & {Mean (\unit{\gray}RBE)} & {SD (\unit{\gray}RBE)} & \makecell[c]{Range (min -- max)\\(\unit{\gray}RBE)} \\[0.5ex]
\toprule
CTV (\qty{5940}{cc}) & P10 & -0.10  & 0.24 & \numrange{-0.72}{1.16} \\
CTV (\qty{5940}{cc}) & P90 & -0.064 & 0.24 & \numrange{-0.62}{1.06} \\
Optic chiasm         & P95 &  0.18  & 0.26 & \numrange{-0.59}{0.92} \\
Optic nerve (L)      & P95 & -0.21  & 0.16 & \numrange{-0.67}{0.57} \\
Optic nerve (R)      & P95 & -0.18  & 0.28 & \numrange{-0.74}{0.68} \\
\bottomrule
\end{tabular}
\end{center}
\end{table}

\subsection{Comparing PCE responses} \label{subsec:PCEresponseComparison}

\begin{figure}[]
    \centering

    % First row
    \begin{subfigure}{0.31\textwidth}
        \centering
        \includegraphics[width=\linewidth]{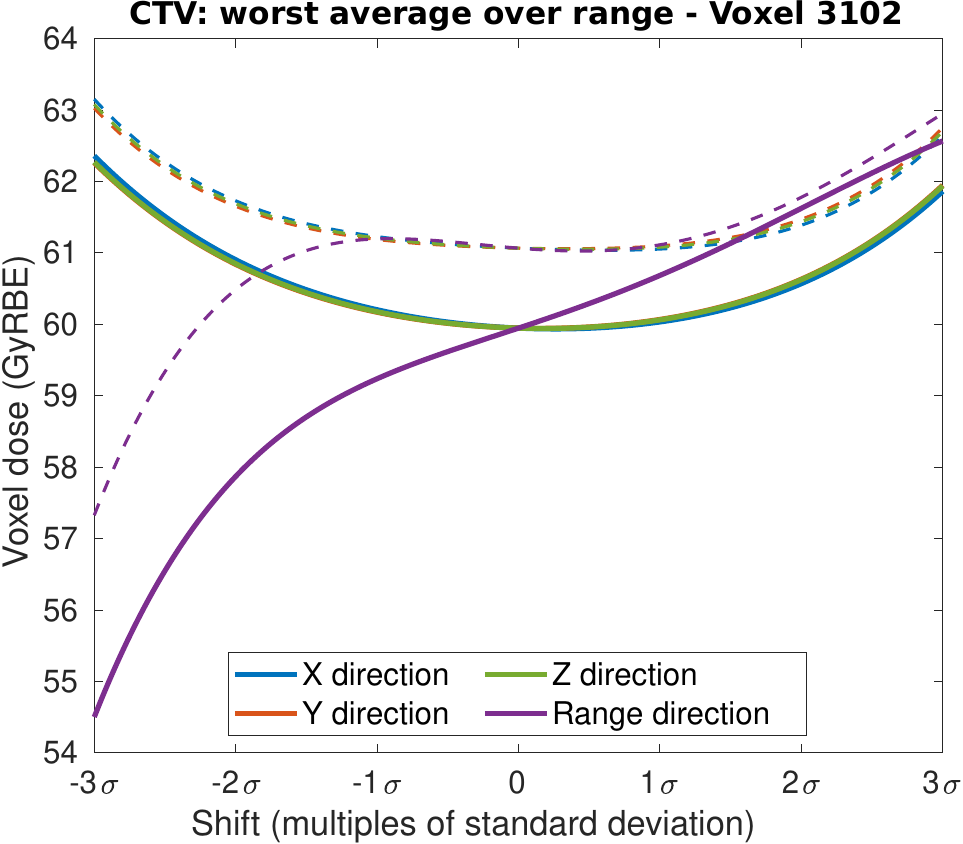}
        \caption{}
    \end{subfigure}
    %\hfill
    \begin{subfigure}{0.31\textwidth}
        \centering
        \includegraphics[width=\linewidth]{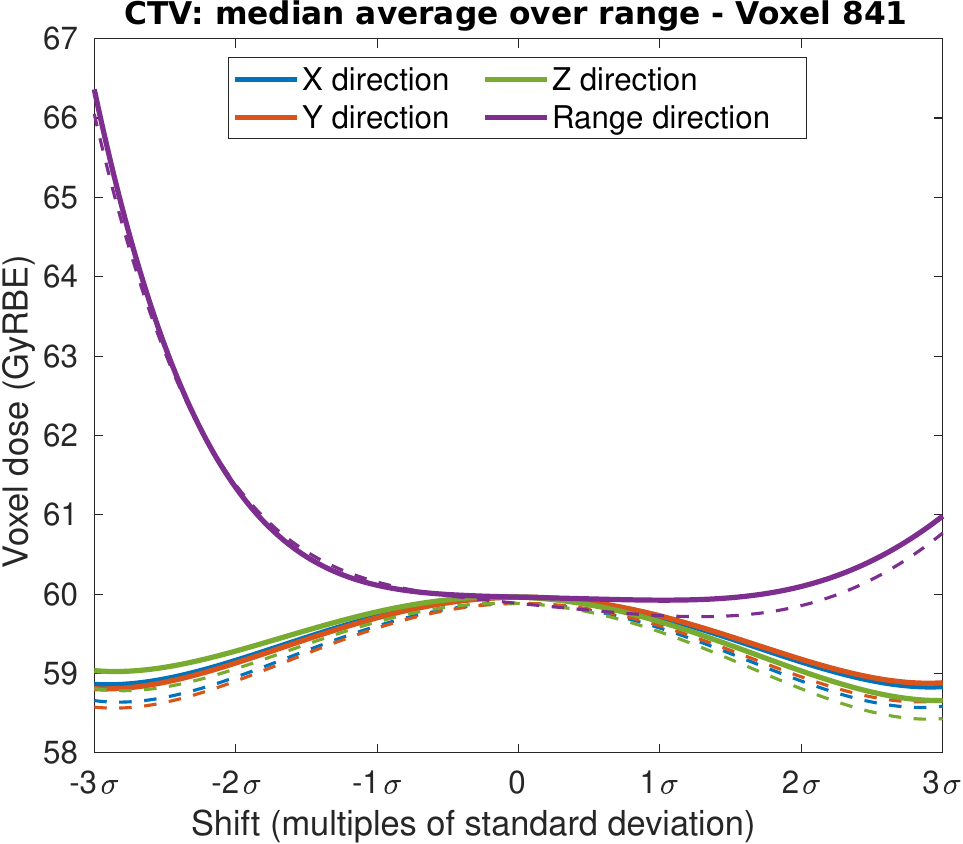}
        \caption{}
    \end{subfigure}
    %\hfill
    \begin{subfigure}{0.31\textwidth}
        \centering
        \includegraphics[width=\linewidth]{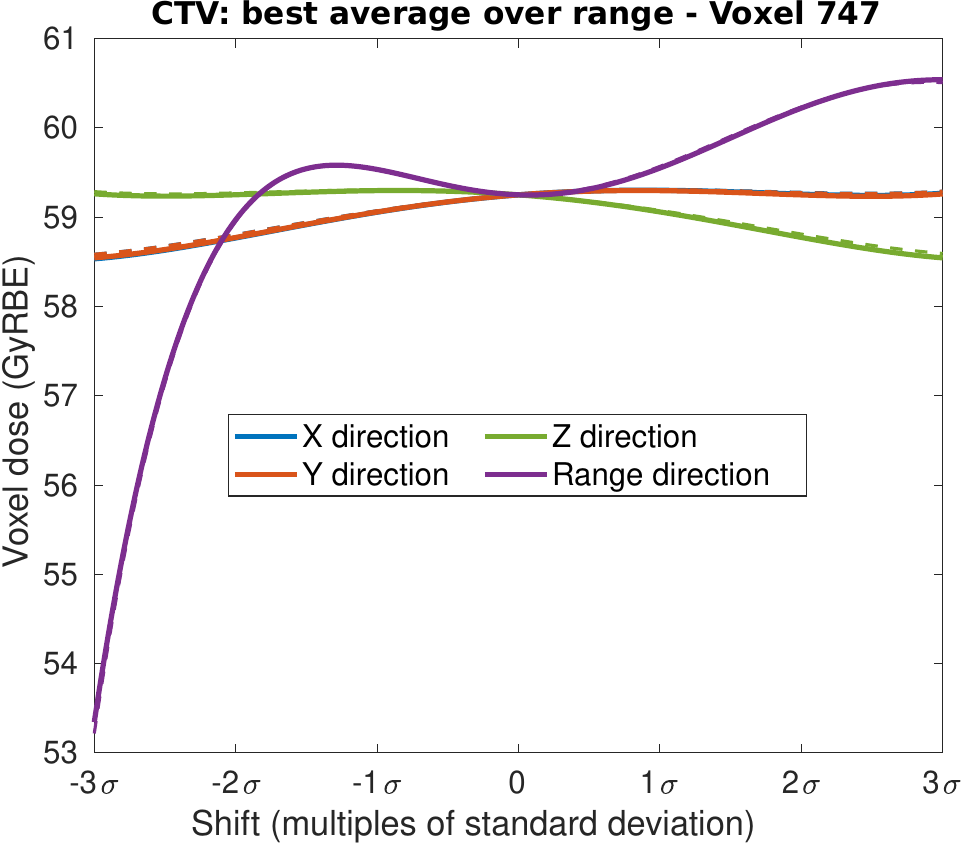}
        \caption{}
    \end{subfigure}

    % Second row
    \begin{subfigure}{0.31\textwidth}
        \centering
        \includegraphics[width=\linewidth]{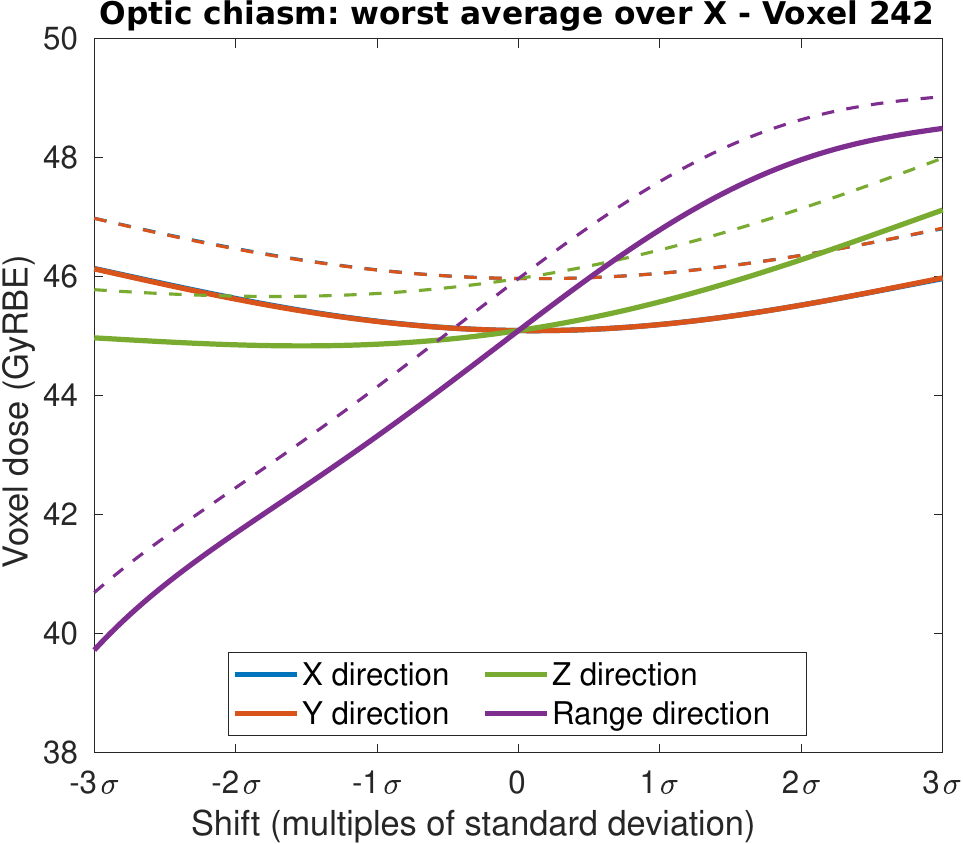}
        \caption{}
    \end{subfigure}
    %\hfill
    \begin{subfigure}{0.31\textwidth}
        \centering
        \includegraphics[width=\linewidth]{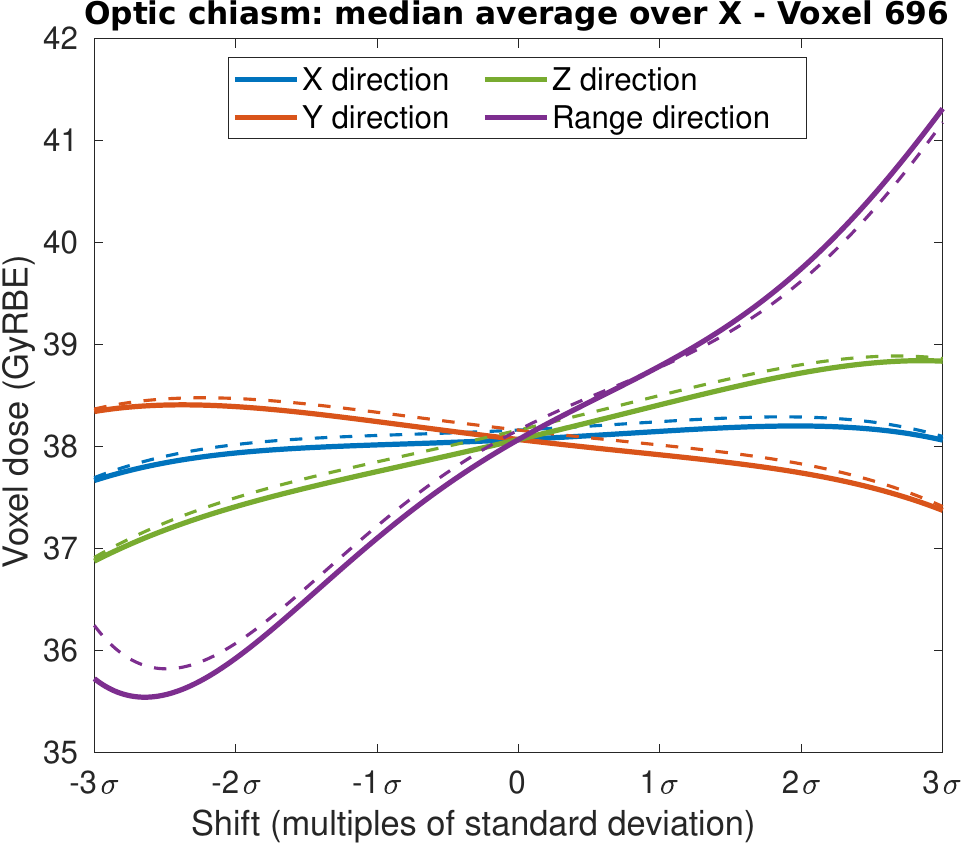}
        \caption{}
    \end{subfigure}
    %\hfill
    \begin{subfigure}{0.31\textwidth}
        \centering
        \includegraphics[width=\linewidth]{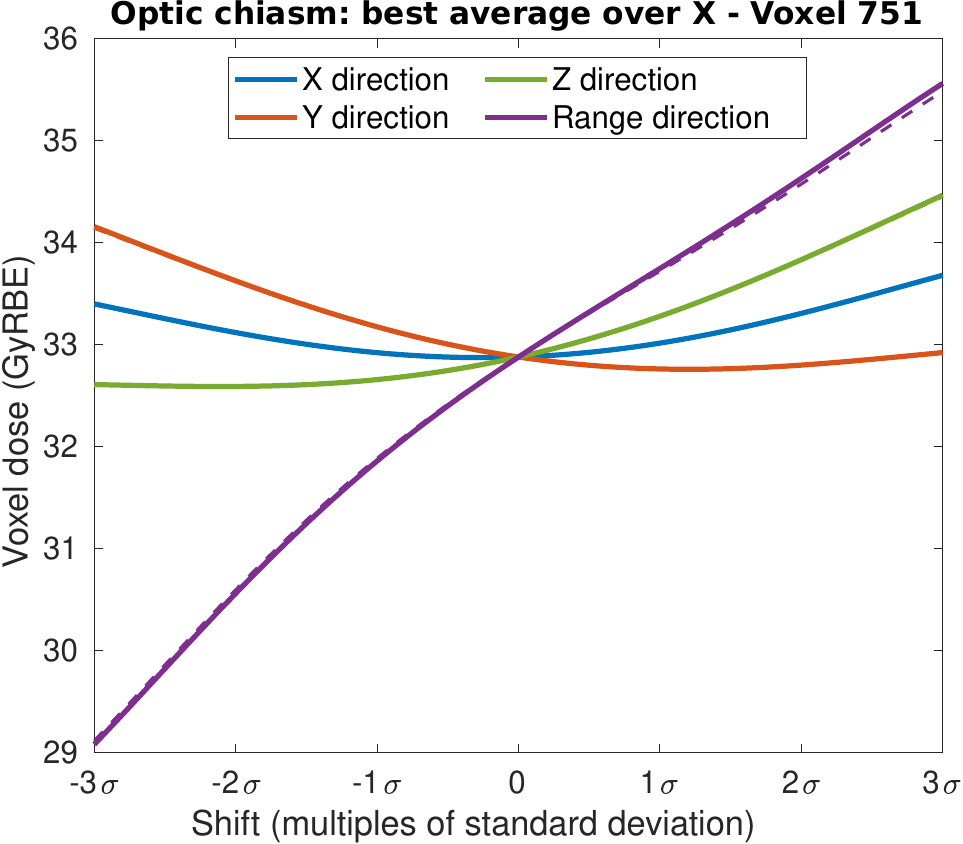}
        \caption{}
    \end{subfigure}

    % Third row
    \begin{subfigure}{0.31\textwidth}
        \centering
        \includegraphics[width=\linewidth]{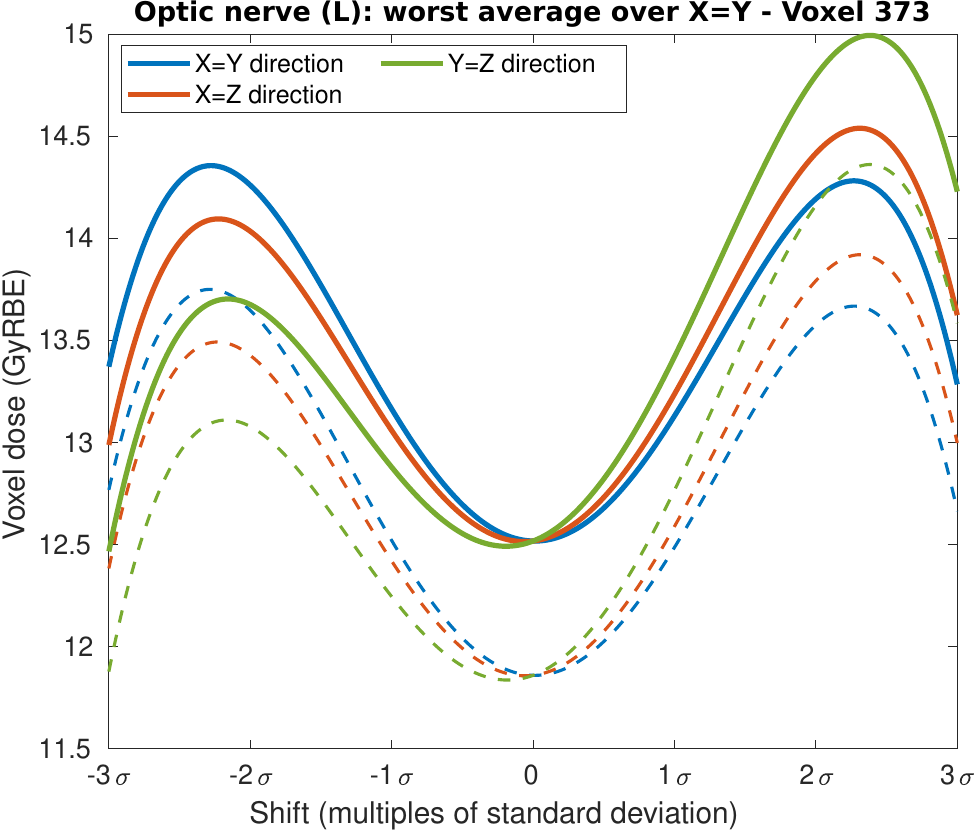}
        \caption{}
    \end{subfigure}
    %\hfill
    \begin{subfigure}{0.31\textwidth}
        \centering
        \includegraphics[width=\linewidth]{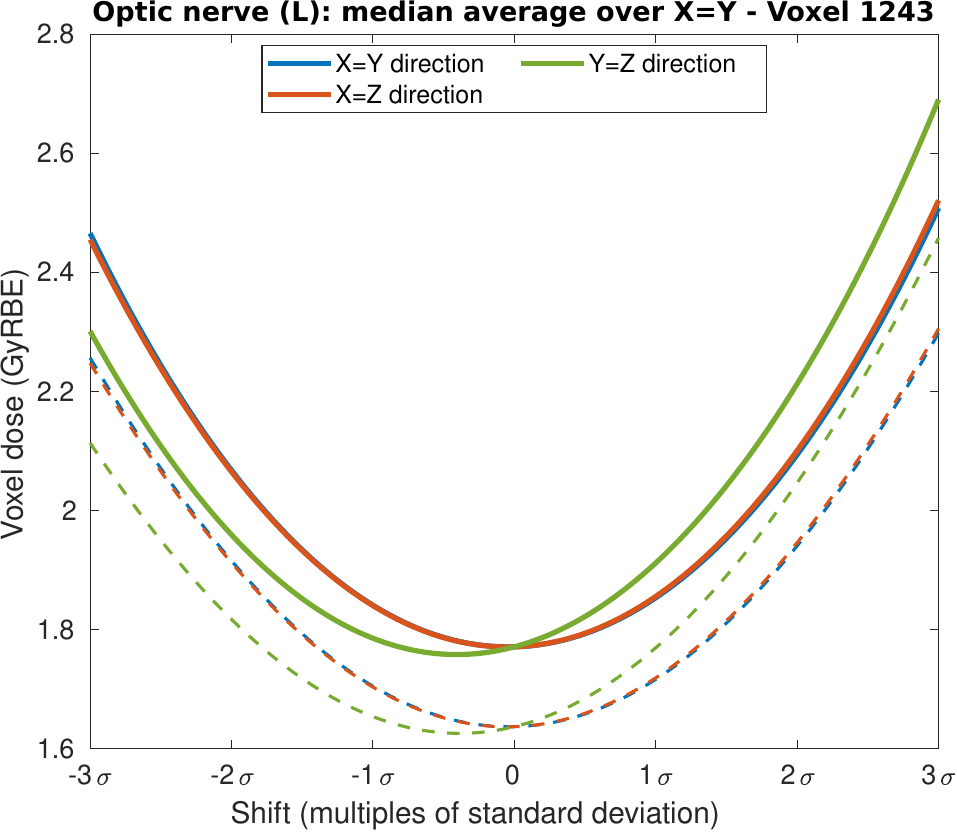}
        \caption{}
    \end{subfigure}
    %\hfill
    \begin{subfigure}{0.31\textwidth}
        \centering
        \includegraphics[width=\linewidth]{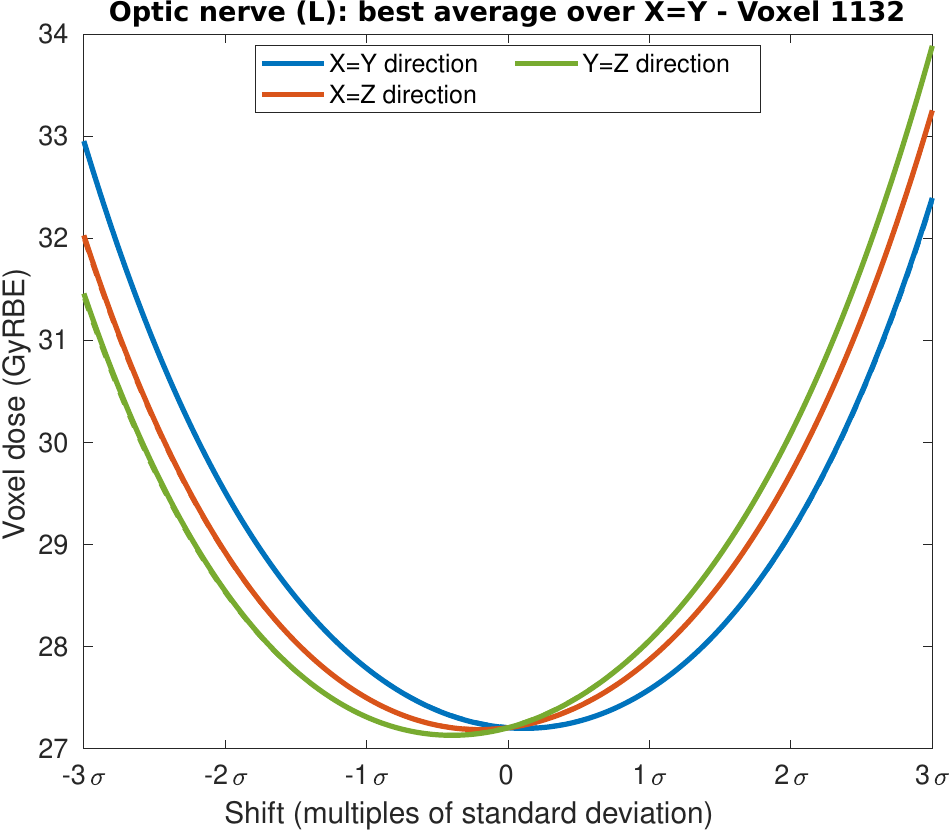}
        \caption{}
    \end{subfigure}

    % Fourth row
    \begin{subfigure}{0.31\textwidth}
        \centering
        \includegraphics[width=\linewidth]{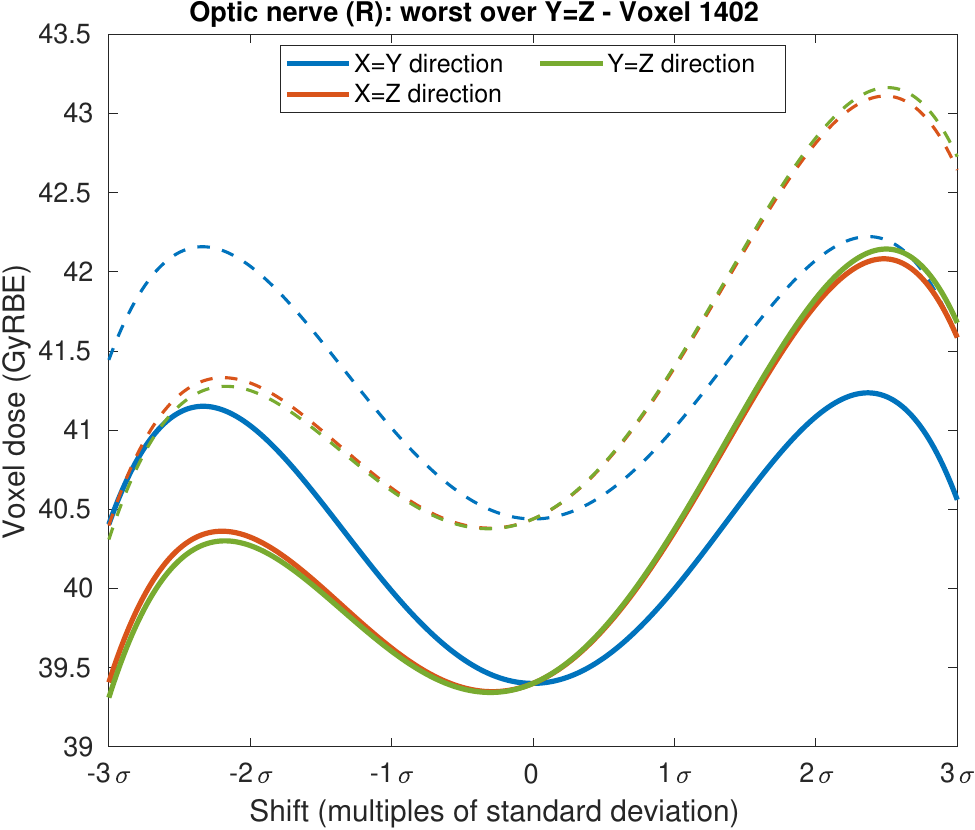}
        \caption{}
    \end{subfigure}
    %\hfill
    \begin{subfigure}{0.31\textwidth}
        \centering
        \includegraphics[width=\linewidth]{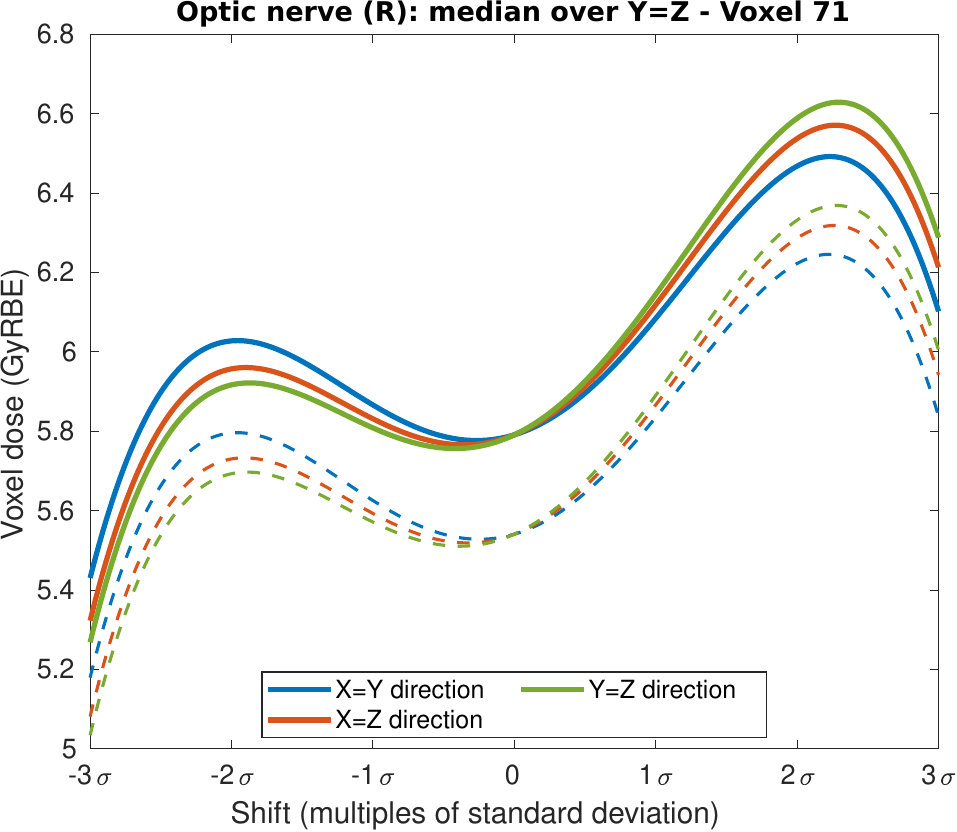}
        \caption{}
    \end{subfigure}
    %\hfill
    \begin{subfigure}{0.31\textwidth}
        \centering
        \includegraphics[width=\linewidth]{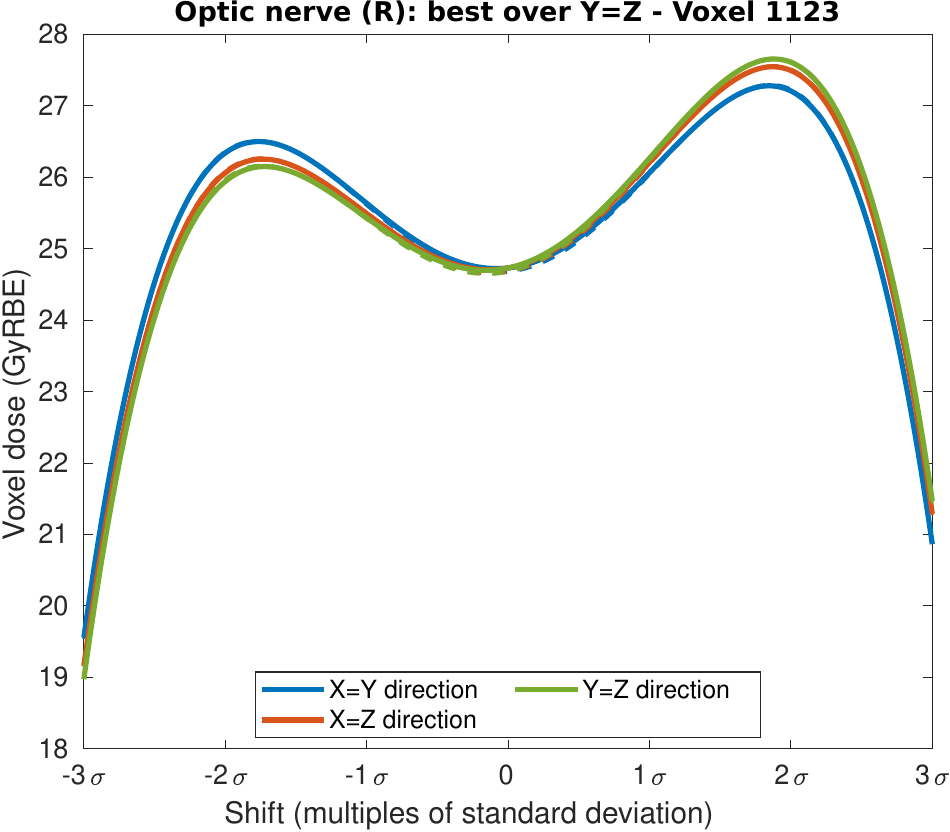}
        \caption{}
    \end{subfigure}

    \caption{A comparison between the $D_{ij} \rightarrow d_i$ (dashed) and voxel dose PCE (solid) for some representative voxels of patient 4 for the CTV and optic system.}
    \label{fig:responseComparison}
\end{figure}

As a visual PCE accuracy validation, we compare some voxel dose distributions in Figure \ref{fig:responseComparison}. The responses are obtained by sampling the $D_{ij} \rightarrow d_i$ and voxel dose PCEs along their principal axes and along the diagonals of the setup errors. We show the worst (1st column), median (2nd column) and best (3rd column) case voxels in terms of mean dose difference along one of the directions, for the CTV (1st row), optic chiasm (2nd row), left optic nerve (3rd row) and right optic nerve (4th row). In most cases the median dose comparison is systematically under- or overdosing the voxel dose PCE, as was also clear from the mean dose difference from Table \ref{tab:ptileDifferences}. Especially clear from the worst case comparisons is that the $D_{ij} \rightarrow d_i$ PCE does not always bias the voxel dose PCE in the same way: the voxel dose PCE does get under- as well as overestimated by the $D_{ij} \rightarrow d_i$ PCE. 

\section{Additional probabilistic plan results}
\subsection{Additional plan comparisons} \label{app:moreResults}
\begin{figure}
    \centering

    % Row 1
    \begin{subfigure}{0.95\textwidth}
        \centering
        \includegraphics[width=\linewidth]{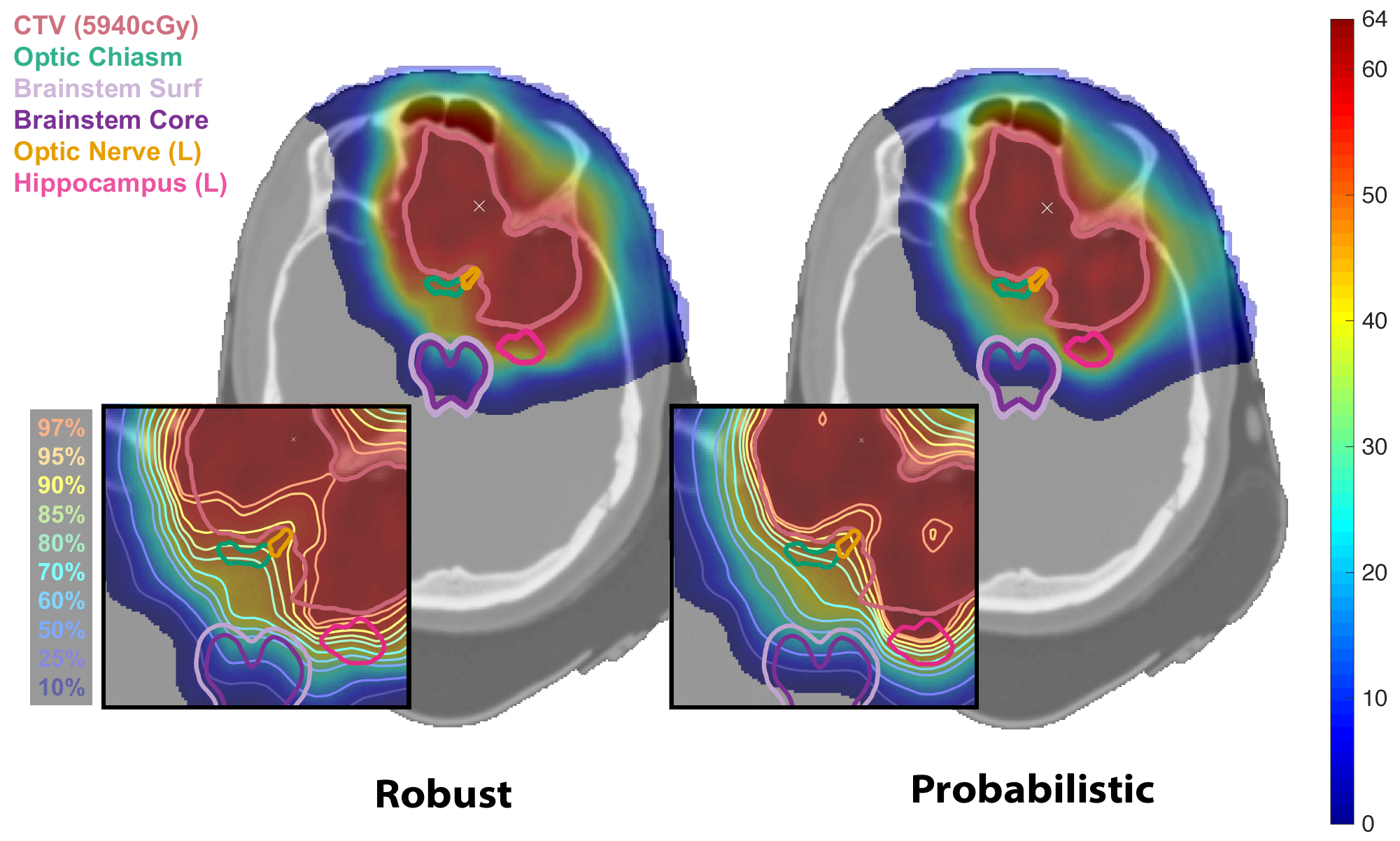}
        \caption{}
        \label{fig:groupB_Neuro10_DPH-CTV}
    \end{subfigure}

    % Row 2
    \begin{subfigure}{0.60\textwidth}
        \centering
        \includegraphics[width=\linewidth]{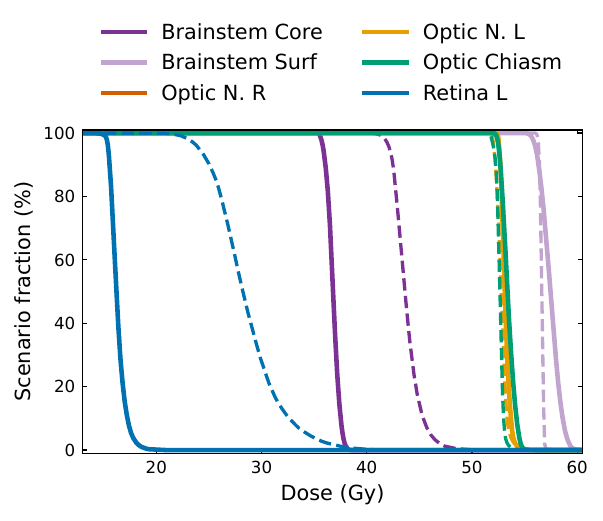}
        \caption{}
    \end{subfigure}

    \caption{A comparison between robust (dashed) and probabilistic (solid) plans for patient 3 in the nominal scenario (a), with the dose population histogram (b) of the $D_{2\%}$ of some relevant OARs. Dose values are in \unit{\gray}RBE.}
    \label{fig:groupB_Neuro12}
\end{figure}

\begin{figure}
    %\centering
    %\includegraphics[width=0.95\linewidth]{Figures/nominal/Neuro32.pdf}
    %\caption{Nominal plan comparison for patient 4 (sagittal plane). Dose values are in \unit{\gray}RBE.}
    %\label{fig:groupB_Neuro32}

    \begin{subfigure}{\textwidth}
        \centering
        \includegraphics[width=0.9\linewidth]{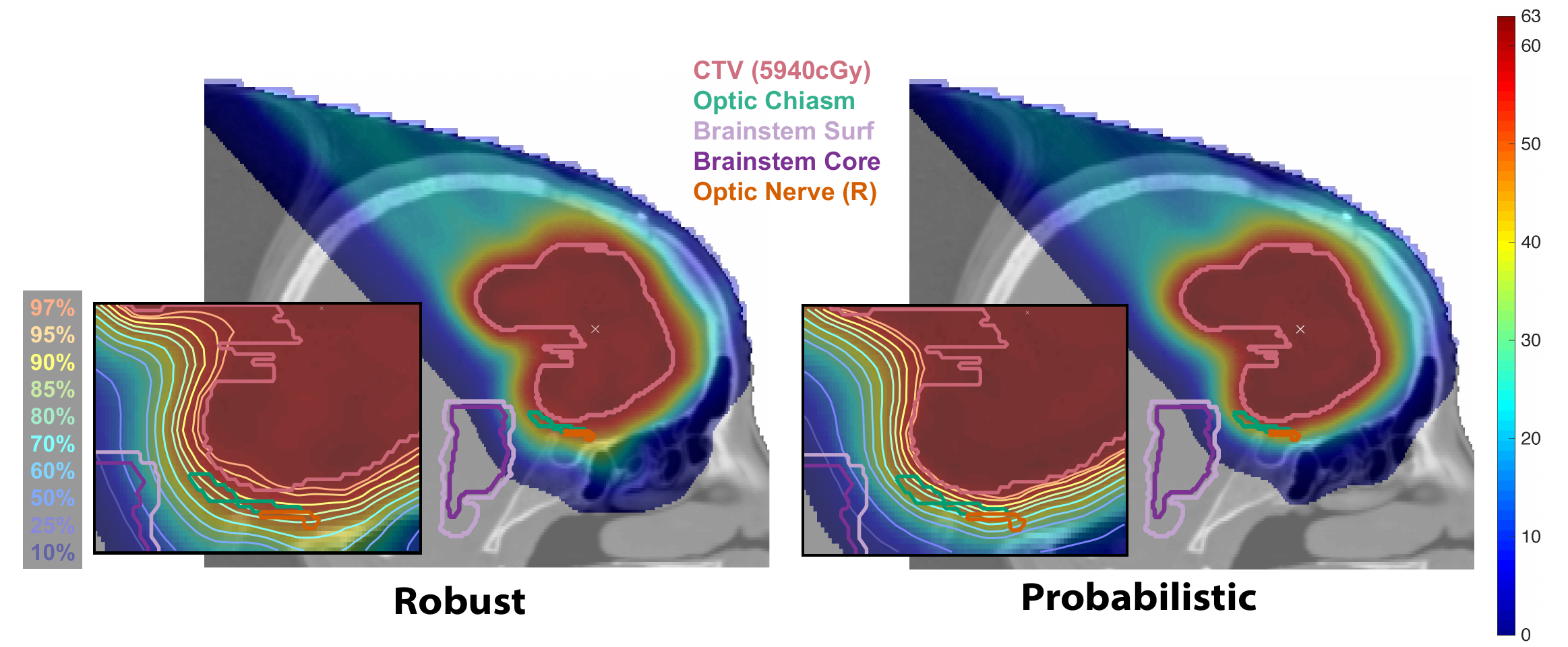}
        \caption{}
    \end{subfigure}
    \begin{subfigure}{\textwidth}
        \centering
        \includegraphics[width=0.9\linewidth]{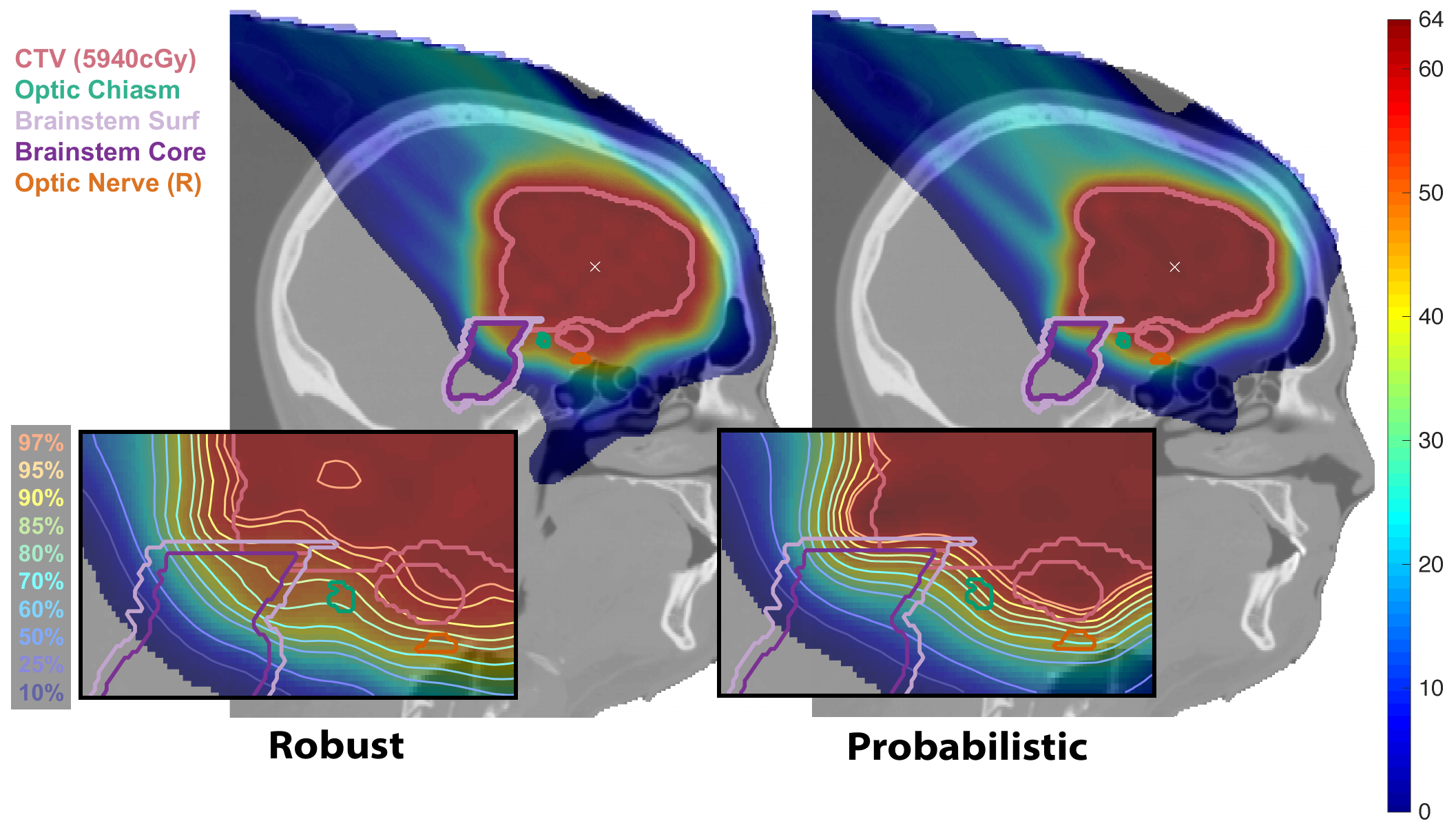}
        \caption{}
    \end{subfigure}
    
    \caption{A comparison between robust and probabilistic plans for a) patient 4 and b) patient 5 in the nominal scenario (sagittal view). Dose values are in \unit{\gray}RBE.}
    \label{fig:groupB_Neuro3235}
\end{figure}

For completeness, Figure \ref{fig:groupB_Neuro12} shows the comparison between the robust and probabilistic plan result in the nominal scenario for patient 3. The dose population histogram (DPH) of the $D_{2\%}$ for relevant OARs is shown as well. Although the $D_{0.03\mathrm{cc},\text{Br.Core}}$ is similar for the robust and probabilistic plan (see Figure \ref{fig:DPHs_groupB_Neuro12OAR}), the $D_{2\%,\text{Br.Core}}$ shows a significant reduction (the 95th percentile reduced by \qty{8.33}{\gray}RBE) for all acceptance probabilities (or scenario fractions), because of the steep dose fall-off in the probabilistic plan. Dose levels of the probabilistic plan are larger in the left hippocampus (with objective weights obtained from Erasmus-iCycle), but can be reduced by tuning its objective weight. Figure \ref{fig:groupB_Neuro3235} shows the nominal plan comparison for patients 4 and 5.

% Neuro32_v15

\begin{table}
\footnotesize
\caption{\label{tab:wishlist_patient4}The wish-list for the robust (Erasmus-iCycle) and probabilistic optimization for patient 1 (with $d_p = \qty{50.4}{\gray}$RBE). Objective/constraint types are either quadratic underdosage penalty (QUP), linear (maximum), mean, or based on voxel dose percentile (e.g., P10 is the 10th percentile of the voxel dose). The objectives in robust planning are priority-based with hard constraints (C). Probabilistic objective weights (OW) are obtained from Erasmus-iCycle.}
\begin{center}
\begin{tabular}{@{}lcllccll}
\toprule
 & \multicolumn{4}{c}{Robust} & \multicolumn{3}{c}{Probabilistic} \\
\cmidrule(lr){2-5} \cmidrule(lr){6-8}
Structure & Priority & Type & \makecell[l]{Goal\\(\unit{\gray}RBE)} & Robust & Type & \makecell[l]{Goal\\(\unit{\gray}RBE)} & OW \\[0.5ex]
\toprule
CTV & C & linear & $97\% \, d_p$ & yes & P10 & $97\% \, d_p$ & C \\
CTV & C & linear & 53.93 & yes & P90 & 53.93 & C \\
CTV (inner ring \qtyrange[range-units=single,range-phrase=--]{0}{2}{\mm}) & C & linear & $97\% \, d_p$ & yes & P10 & $97\% \, d_p$ & C \\
CTV (inner ring \qtyrange[range-units=single,range-phrase=--]{0}{2}{\mm}) & C & linear & 53.93 & yes & P90 & 53.93 & C \\
External & C & linear & 52.8 & -- & & 52.8 & C \\
\midrule
CTV (ring \qtyrange[range-units=single,range-phrase=--]{2}{5}{\mm})   & 1  & linear & $90\% \, d_p$ & -- & & $90\% \, d_p$ & \num{2.46e-2} \\
CTV (ring \qtyrange[range-units=single,range-phrase=--]{5}{7}{\mm})   & 1  & linear & $60\% \, d_p$ & -- & & $60\% \, d_p$ & \num{8.23e-2} \\
CTV (ring \qtyrange[range-units=single,range-phrase=--]{7}{10}{\mm})  & 1  & linear & $30\% \, d_p$ & -- & & $30\% \, d_p$ & \num{0.185}   \\
Brainstem Surf  & 2  & linear & 32   & -- & & 32   & \num{0.439}   \\
Optic Chiasm    & 3  & linear & 45   & -- & & 45   & \num{1.82e-6} \\
Optic Nerve (L) & 3  & linear & 47   & -- & & 47   & \num{5.66e-2} \\
Optic Nerve (R) & 3  & linear & 11   & -- & & 11   & \num{7.46e-2} \\
Retina (L)      & 4  & linear & 11   & -- & & 11   & \num{3.42e-5} \\
Pituitary       & 5  & mean   & 17   & -- & & 17   & \num{0.128}   \\
Hippocampus (L) & 6  & mean   & 14.5 & -- & & 14.5 & \num{1.47e-8} \\
Brain-CTV       & 7  & mean   & 4.5  & -- & & 4.5  & \num{1.09e-7} \\
Cochlea (L)     & 8  & mean   & 0.5  & -- & & 0.5  & \num{1.07e-8} \\
External        & 9  & linear & 0    & -- & & 0    & \num{5.80e-3} \\
MU              & 10 & linear & 1    & -- & &      & \num{2.60e-3} \\
\bottomrule
\end{tabular}
\end{center}
\end{table}

\begin{table}
\footnotesize
\caption{\label{tab:wishlist_patient10}The wish-list for the robust (Erasmus-iCycle) and probabilistic optimization for patient 2 (with $d_p = \qty{59.4}{\gray}$RBE). Objective/constraint types are either quadratic underdosage penalty (QUP), linear (maximum), mean, or based on voxel dose percentile (e.g., P10 is the 10th percentile of the voxel dose). The objectives in robust planning are priority-based with hard constraints (C). Probabilistic objective weights (OW) are obtained from Erasmus-iCycle.}
\begin{center}
\begin{tabular}{@{}lcllccll}
\toprule
 & \multicolumn{4}{c}{Robust} & \multicolumn{3}{c}{Probabilistic} \\
\cmidrule(lr){2-5} \cmidrule(lr){6-8}
Structure & Priority & Type & \makecell[l]{Goal\\(\unit{\gray}RBE)} & Robust & Type & \makecell[l]{Goal\\(\unit{\gray}RBE)} & OW \\[0.5ex]
\toprule
CTV                          & C & linear & $107\% \, d_p$ & yes & P90 & $107\% \, d_p$ & C \\
External                     & C & linear & 63.26 & --  &     & 63.26 & C \\
Brainstem Surface            & C & linear & 60    & yes & P95 & 59.92 & C \\
Brainstem Core               & C & linear & 53.7  & yes & P95 & 53.3  & C \\
CTV and Brainstem Core       & C & linear & 53.7  & yes & P95 & 53.3  & C \\
CTV and Brainstem Surf       & C & linear & 60    & yes & P95 & 59.92 & C \\
Brainstem Core close to CTV  & C & linear & 53.7  & yes & P95 & 53.3  & C \\
\midrule
CTV                          & 1 & QUP & $99\% \, d_p$ & yes & P10 & $100\% \, d_p$ & \num{0.091}   \\
CTV and Brainstem Core       & 1 & QUP & $99\% \, d_p$ & yes & P10 & $100\% \, d_p$ & \num{6.9e-2}  \\
CTV and Brainstem Surf       & 1 & QUP & $99\% \, d_p$ & yes & P10 & $100\% \, d_p$ & \num{3.7e-2}  \\
CTV close to Brainstem Surf  & 1 & QUP & $99\% \, d_p$ & yes & P10 & $100\% \, d_p$ & \num{6.7e-2}  \\
CTV (inner ring \qtyrange[range-units=single,range-phrase=--]{0}{2}{\mm}) & 1 & QUP & $99\% \, d_p$ & yes & P10 & $100\% \, d_p$ & \num{0.72} \\
CTV (ring \qtyrange[range-units=single,range-phrase=--]{2}{5}{\mm})   & 2  & linear & $80\% \, d_p$ & -- & & $80\% \, d_p$ & \num{1.6e-3}  \\
CTV (ring \qtyrange[range-units=single,range-phrase=--]{5}{7}{\mm})   & 2  & linear & $65\% \, d_p$ & -- & & $65\% \, d_p$ & \num{4.3e-3}  \\
CTV (ring \qtyrange[range-units=single,range-phrase=--]{7}{10}{\mm})  & 2  & linear & $50\% \, d_p$ & -- & & $50\% \, d_p$ & \num{5.0e-3}  \\
Optic Chiasm       & 3  & linear & 42 & -- & & 42 & \num{6.0e-3}   \\
Optic Nerve (L)    & 4  & linear & 39 & -- & & 39 & \num{6.1e-4}   \\
skin tpv CTV       & 5  & linear & 50 & -- & & 50 & \num{4.0e-4}   \\
Hippocampus (R)    & 6  & mean   & 5  & -- & & 5  & \num{2.1e-3}   \\
Pituitary          & 7  & mean   & 10 & -- & & 10 & \num{3.0e-10}  \\
Cochlea (L)        & 8  & mean   & 39 & -- & & 39 & \num{8.4e-11}  \\
Glnd Lacrimal (L)  & 9  & mean   & 7  & -- & & 7  & \num{6.7e-11}  \\
Brain-CTV          & 10 & mean   & 10 & -- & & 10 & \num{3.1e-4}   \\
External           & 11 & linear & 0  & -- & & 0  & \num{1.6e-4}   \\
MU                 & 12 & linear & 1  & -- & & & \num{1.2e-5}      \\
\bottomrule
\end{tabular}
\end{center}
\end{table}

\begin{table}
\footnotesize
\caption{\label{tab:wishlist_patient12}The wish-list for the robust (Erasmus-iCycle) and probabilistic optimization for patient 3 (with $d_p = \qty{59.4}{\gray}$RBE). Objective/constraint types are either quadratic underdosage penalty (QUP), linear (maximum), mean, or based on voxel dose percentile (e.g., P10 is the 10th percentile of the voxel dose). The objectives in robust planning are priority-based with hard constraints (C). Probabilistic objective weights (OW) are obtained from Erasmus-iCycle.}
\begin{center}
\begin{tabular}{@{}lcllccll}
\toprule
 & \multicolumn{4}{c}{Robust} & \multicolumn{3}{c}{Probabilistic} \\
\cmidrule(lr){2-5} \cmidrule(lr){6-8}
Structure & Priority & Type & \makecell[l]{Goal\\(\unit{\gray}RBE)} & Robust & Type & \makecell[l]{Goal\\(\unit{\gray}RBE)} & OW \\[0.5ex]
\toprule
CTV                     & C & linear & $107\% \, d_p$ & yes & P90 & $107\% \, d_p$ & C \\
Optic Chiasm            & C & linear & 55             & yes & P95 & 54.69          & C \\
Optic Nerve (R)         & C & linear & 55             & yes & P95 & 55             & C \\
Optic Nerve (L)         & C & linear & 55             & yes & P95 & 54.65          & C \\
Brainstem Core          & C & linear & 54             & yes & P95 & 52.43          & C \\
Brainstem Surface       & C & linear & 60             & yes & P95 & 59.71          & C \\
Retina (L)              & C & linear & 46             & yes & P95 & 46             & C \\
CTV and Brainstem Surf  & C & linear & 60             & yes & P95 & 59.71          & C \\
External                & C & linear & $104\% \, d_p$ & --  &     & $104\% \, d_p$ & C \\
\midrule
CTV                          & 1 & QUP & $99\% \, d_p$ & yes & P10 & $99\% \, d_p$ & \num{0.1198} \\
CTV and Brainstem Surf       & 1 & QUP & $99\% \, d_p$ & yes & P10 & $99\% \, d_p$ & \num{0.206}  \\
CTV close to Brainstem Surf  & 1 & QUP & $99\% \, d_p$ & yes & P10 & $99\% \, d_p$ & \num{0.268}  \\
CTV (inner ring \qtyrange[range-units=single,range-phrase=--]{0}{2}{\mm}) & 1 & QUP & $99\% \, d_p$ & yes & P10 & $99\% \, d_p$ & \num{0.339} \\
CTV (ring \qtyrange[range-units=single,range-phrase=--]{2}{5}{\mm})   & 2  & linear & $80\% \, d_p$ & -- & & $80\% \, d_p$ & \num{2.10e-3} \\
CTV (ring \qtyrange[range-units=single,range-phrase=--]{5}{7}{\mm})   & 2  & linear & $65\% \, d_p$ & -- & & $65\% \, d_p$ & \num{3.70e-3} \\
CTV (ring \qtyrange[range-units=single,range-phrase=--]{7}{10}{\mm})  & 2  & linear & $50\% \, d_p$ & -- & & $50\% \, d_p$ & \num{4.50e-3} \\
Hippocampus (R)   & 3  & mean   & 0.5  & -- & & 0.5  & \num{5.28e-2} \\
Hippocampus (L)   & 4  & linear & 62   & -- & & 62   & \num{2.16e-4} \\
Pituitary         & 5  & mean   & 34   & -- & & 34   & \num{1.7e-3}  \\
Retina (R)        & 6  & mean   & 10   & -- & & 10   & \num{9.97e-9} \\
Brainstem         & 6  & mean   & 10   & -- & & 10   & \num{6.11e-4} \\
Lens (R)          & 7  & mean   & 3    & -- & & 3    & \num{4.26e-5} \\
Lens (L)          & 7  & mean   & 5    & -- & & 5    & \num{6.53e-9} \\
Cochlea (L)       & 8  & mean   & 5    & -- & & 5    & \num{3.18e-6} \\
Glnd Lacrimal (L) & 9  & mean   & 25.5 & -- & & 25.5 & \num{3.54e-8} \\
Brain-CTV         & 10 & mean   & 10   & -- & & 10   & \num{1.7e-3}  \\
External          & 11 & linear & 0    & -- & & 0    & \num{4.93e-4} \\
MU                & 12 & linear & 1    & -- & &      & \num{1.44e-6} \\
\bottomrule
\end{tabular}
\end{center}
\end{table}

\begin{table}
\footnotesize
\caption{\label{tab:wishlist_patient32}The wish-list for the robust (Erasmus-iCycle) and probabilistic optimization for patient 4 (with $d_p = \qty{59.4}{\gray}$RBE). Objective/constraint types are either quadratic underdosage penalty (QUP), linear (maximum), mean, or based on voxel dose percentile (e.g., P10 is the 10th percentile of the voxel dose). The objectives in robust planning are priority-based with hard constraints (C). Probabilistic objective weights (OW) are obtained from Erasmus-iCycle.}
\begin{center}
\begin{tabular}{@{}lcllccll}
\toprule
 & \multicolumn{4}{c}{Robust} & \multicolumn{3}{c}{Probabilistic} \\
\cmidrule(lr){2-5} \cmidrule(lr){6-8}
Structure & Priority & Type & \makecell[l]{Goal\\(\unit{\gray}RBE)} & Robust & Type & \makecell[l]{Goal\\(\unit{\gray}RBE)} & OW \\[0.5ex]
\toprule
CTV             & C & linear & $107\% \, d_p$ & yes & P90 & $105\% \, d_p$ & C \\
Optic Chiasm    & C & linear & 55.5           & yes & P95 & 55.5           & C \\
Optic Nerve (R) & C & linear & 55             & yes & P95 & 55             & C \\
Optic Nerve (L) & C & linear & 55             & yes & P95 & 55             & C \\
External        & C & linear & $105\% \, d_p$ & --  &     & $105\% \, d_p$ & C \\
\midrule
CTV & 1 & QUP & $97\% \, d_p$ & yes & P10 & 58.8 & \num{0.9955} \\
CTV (ring \qtyrange[range-units=single,range-phrase=--]{2}{5}{\mm})   & 2 & linear & $80\% \, d_p$ & -- & & $80\% \, d_p$ & \num{1.08e-4} \\
CTV (ring \qtyrange[range-units=single,range-phrase=--]{5}{7}{\mm})   & 2 & linear & $65\% \, d_p$ & -- & & $65\% \, d_p$ & \num{8.43e-4} \\
CTV (ring \qtyrange[range-units=single,range-phrase=--]{7}{10}{\mm})  & 2 & linear & $50\% \, d_p$ & -- & & $50\% \, d_p$ & \num{1.8e-3}  \\
Glnd Lacrimal (R) & 3 & mean   & 8  & -- & & 8  & \num{3.23e-13} \\
Brain-CTV         & 4 & mean   & 11 & -- & & 11 & \num{5.79e-13} \\
External          & 5 & linear & 0  & -- & & 0  & \num{1.7e-3}   \\
MU                & 6 & linear & 1  & -- & &    & \num{1.31e-5}  \\
\bottomrule
\end{tabular}
\end{center}
\end{table}

\begin{table}
\footnotesize
\caption{\label{tab:wishlist_patient35}The wish-list for the robust (Erasmus-iCycle) and probabilistic optimization for patient 5 (with $d_p = \qty{59.4}{\gray}$RBE). Objective/constraint types are either quadratic underdosage penalty (QUP), linear (maximum), mean, or based on voxel dose percentile (e.g., P10 is the 10th percentile of the voxel dose). The objectives in robust planning are priority-based with hard constraints (C). Probabilistic objective weights (OW) are obtained from Erasmus-iCycle.}
\begin{center}
\begin{tabular}{@{}lcllccll}
\toprule
 & \multicolumn{4}{c}{Robust} & \multicolumn{3}{c}{Probabilistic} \\
\cmidrule(lr){2-5} \cmidrule(lr){6-8}
Structure & Priority & Type & \makecell[l]{Goal\\(\unit{\gray}RBE)} & Robust & Type & \makecell[l]{Goal\\(\unit{\gray}RBE)} & OW \\[0.5ex]
\toprule
CTV                    & C & linear & $107\% \, d_p$ & yes & P90 & $107\% \, d_p$ & C \\
Optic Chiasm           & C & linear & 55             & yes & P95 & 54.83          & C \\
Optic Nerve (R)        & C & linear & 55             & yes & P95 & 55             & C \\
Brainstem Core         & C & linear & 53.8           & yes & P95 & 51.7           & C \\
Brainstem Surface      & C & linear & 60             & yes & P95 & 60             & C \\
CTV and Brainstem Surf & C & linear & 60             & yes & P95 & 60             & C \\
External               & C & linear & 62             & --  &     & 62             & C \\
\midrule
CTV                     & 1 & QUP    & $99\% \, d_p$ & yes & P10 & $100\% \, d_p$ & \num{0.2377} \\
CTV and Brainstem Surf  & 1 & QUP    & $99\% \, d_p$ & yes & P10 & $100\% \, d_p$ & \num{0.0815} \\
CTV (inner ring \qtyrange[range-units=single,range-phrase=--]{0}{2}{\mm}) & 1 & QUP & $99\% \, d_p$ & yes & P10 & $100\% \, d_p$ & \num{0.573} \\
CTV                     & 1 & linear & 54            & yes & P10 & 55             & \num{3e-3}   \\
CTV (ring \qtyrange[range-units=single,range-phrase=--]{2}{5}{\mm})   & 2 & linear & $90\% \, d_p$ & -- & & $90\% \, d_p$ & \num{0.0017} \\
CTV (ring \qtyrange[range-units=single,range-phrase=--]{5}{7}{\mm})   & 2 & linear & $60\% \, d_p$ & -- & & $60\% \, d_p$ & \num{0.0062} \\
CTV (ring \qtyrange[range-units=single,range-phrase=--]{7}{10}{\mm})  & 2 & linear & $30\% \, d_p$ & -- & & $30\% \, d_p$ & \num{0.0097} \\
Brainstem Surface & 3 & $D_{1\mathrm{cc}}$    & 43   & -- & & 43   & \num{0.0354}    \\
Brainstem Core    & 3 & $D_{0.70\mathrm{cc}}$ & 40   & -- & & 40   & \num{0.031}     \\
Skin              & 4 & linear                & 50   & -- & & 50   & \num{0.0031}    \\
Hippocampus (R)   & 5 & mean                  & 21.5 & -- & & 21.5 & \num{3.7349e-4} \\
Pituitary         & 6 & mean                  & 7.2  & -- & & 7.2  & \num{0.0048}    \\
Eye (R)           & 6 & mean                  & 5    & -- & & 5    & \num{8.3065e-10}\\
Glnd Lacrimal (R) & 7 & mean                  & 24   & -- & & 24   & \num{0.0012}    \\
Brain-CTV         & 8 & mean                  & 13   & -- & & 13   & \num{1.13e-2}   \\
MU                & 9 & linear                & 1    & -- & &      & \num{1.44e-6}   \\
\bottomrule
\end{tabular}
\end{center}
\end{table}

\subsection{Patient-specific wish-lists} \label{app:wishlists}
Wish-lists for patients 1 to 5 are shown in Table \ref{tab:wishlist_patient4} to Table \ref{tab:wishlist_patient35}, respectively. Robust optimization is priority-based, meaning that the first optimization is optimizing for the objectives with priority 1, subject to all hard constraints. Once the optimal solution is found, the obtained dose thresholds associated to the just optimized objectives are relaxed by 3\% and reformulated as hard constraints in the next iteration. This procedure continues until all objectives have been optimized for sequentially. Nominal objectives and constraints (i.e., the non-robust ones) stay nominal. All robust objectives and constraints in the wish-list are converted into probabilistic terms. For the probabilistic objectives, objective weights were obtained from the Lagrange multipliers corresponding to the Erasmus-iCycle plan (see Section \ref{subsec:probWeights} of the manuscript). The wish-list is adapted from the previously defined objectives and constraints from Raystation (previously used in \citet{deJong2025}). Although patient-specific wish-lists were used in this work, they could be adapted such that a single wish-list can be used for the neuro-oncological patient cohort.

\newpage
\thispagestyle{empty}
\mbox{}
\newpage

\subsection{Criteria on percentile convergence} \label{app:convCriteria}
In the following, more details are given on configurations for percentile convergence. Probabilistic optimizations were done using different configurations to show a variety of possibilities.

\subsubsection{Convergence by moving average}
If moving averages are used for percentile convergence, we require all \textit{undamped} voxel dose percentiles to satisfy a convergence criterion that is based on the stability of their trends. The trend is quantified by smoothing the \textit{undamped} percentiles at iteration $k$ using moving average $m_k$ with a window size $W$. For a lag of $\Delta K$ iterations and convergence tolerance $\tau_i$, the outer loop is considered converged if for all voxels $i$,
\begin{equation} \label{eq:MA}
    |m_{i}^{k} - m_i^{k-\Delta K}| < \tau_i,
\end{equation}
where 
\begin{equation} \label{eq:convTol_MA}
    \tau_i = \left\{
    \begin{array}{ll}
        \tau_i^{\mathrm{rel}} \cdot |m_{i}^{k}| & \mbox{if } |m_{i}^{k}| > d_{\mathrm{low}} \\
        \tau_i^{\mathrm{abs}}                   & \mbox{otherwise.}
    \end{array}
    \right.
\end{equation}
For some cases we distinguish between a relative ($\tau_i^{\text{rel}}$) and absolute ($\tau_i^{\text{abs}}$) tolerance, because in low-dose regions ($< d_{\text{low}}$) relative dose differences could become large for low $|m_{i}^{k}|$. Unless specified differently, $\tau_i^{\text{rel}} = 0.01$, $\tau_i^{\text{abs}} = \qty{1}{\gray}\text{RBE}$ and $d_{\text{low}} = \qty{5}{\gray}\text{RBE}$.

\subsubsection{Convergence by relative $L_2$-norm}
Percentile convergence using moving averages can become overly strict, because Equation \ref{eq:MA} has to be met for all voxels. Also, it involves smoothing with window $W$ and uses a lag $\Delta K$, meaning that at least $W + \Delta K$ outer iterations have to be done before the convergence criteria can be calculated.

An alternative convergence metric that does not rely on these is based on the relative $L_2$-norm difference $\Delta \bi{d}^{\alpha \%}_{s}(k)$ between percentiles at iteration $k$ and $k-1$, as
\begin{equation}
    \Delta \bi{d}_s^{\alpha \%}(k) = \frac{\Vert \bi{d}^{\alpha \%}_{k,s} - \bi{d}^{\alpha \%}_{k-1,s} \Vert}{\Vert \bi{d}^{\alpha \%}_{k,s} \Vert}
\end{equation}
for structure $s$. The optimization is converged  if $\Delta \bi{d}^{\alpha \%}_{s}(k) < \epsilon_s, \forall s \in S_{prob}$, where convergence threshold $\epsilon_s$ is possibly structure-dependent.

\begin{table}
\caption{\label{tab:probConvSettings}Overview of convergence settings for the probabilistic plans. Plans were either optimized by a single phase (1P) with reduced tolerances (RT) or two-phase (2P) optimization, where phase-II used convergence on moving average $\text{MA}(W,\Delta K)$ for $W = 3$ and $\Delta K = 3$.}
\begin{center}
\begin{tabular}{@{}ccccclc}
\toprule
Pt & Conv.\ method & $\tau_i^{\text{rel}}$ & $\tau_i^{\text{abs}}$ & $d_{\text{low}}$ & Optimization strategy & Iterations \\[0.5ex]
\toprule
4  & $\text{MA}(3,3)$ & 0.01 & --                & \qty{0}{\gray}RBE & 1P, RT($10^{-2} \rightarrow 10^{-5}$) & 10 \\
10 & $\text{MA}(3,3)$ & 0.01 & --                & \qty{0}{\gray}RBE & 1P, RT($10^{-2} \rightarrow 10^{-5}$) & 25 \\
12 & $\text{MA}(3,3)$ & 0.05 & \qty{1}{\gray}RBE & \qty{5}{\gray}RBE & 2P (switched at it.\ 10)              & 12 \\
32 & $\text{MA}(3,3)$ & 0.05 & \qty{1}{\gray}RBE & \qty{5}{\gray}RBE & 2P (switched at it.\ 6)               & 11 \\
35 & $\text{MA}(3,3)$ & 0.05 & \qty{1}{\gray}RBE & \qty{5}{\gray}RBE & 2P (switched at it.\ 10)              & 20 \\
\bottomrule
\end{tabular}
\end{center}
\end{table}
% Neuro32_v15, row: 32 & $L_2$ & -- & -- & -- & 0.01 & 2P (switched at it. 4) & 7 \\ (column 6 is \epsilon parameter for L2 norm)

\begin{table}[t]
\caption{\label{tab:probConvSettingsPatientSpecific}Patient-specific convergence settings to show its dependence on number of outer iterations. All cases use moving average convergence for window $W = 2$ and lag $\Delta K = 2$, with $\tau_i^{\mathrm{rel}} = 0.01$, $\tau_i^{\mathrm{abs}} = \qty{0.2}{\gray}\text{RBE}$, and $d_{\mathrm{low}} = \qty{5}{\gray}\text{RBE}$. Identical optimization strategies are used as in Table~\ref{tab:probConvSettings}.}
\begin{center}
\begin{tabular}{@{}ccl@{}}
\toprule
Pt & Number outer iterations & Structure-specific tolerances \\[0.5ex]
\toprule
4  & 8  & -- \\
\midrule
10 & 10 & \makecell[l]{Brainstem Core/Surf: $\tau_i^{\text{rel}} = 0.1$, $\tau_i^{\text{abs}} = \qty{0.5}{\gray}\text{RBE}$} \\
\midrule
12 & 10 & \makecell[l]{Brainstem Core/Surf: $\tau_i^{\text{rel}} = 0.1$, $\tau_i^{\text{abs}} = \qty{0.5}{\gray}\text{RBE}$ \\
                       Optic nerves: $\tau_i^{\text{rel}} = 0.025$, $\tau_i^{\text{abs}} = \qty{0.5}{\gray}\text{RBE}$ \\
                       Retina L: $\tau_i^{\text{rel}} = 0.02$, $\tau_i^{\text{abs}} = \qty{0.5}{\gray}\text{RBE}$} \\
\bottomrule
\end{tabular}
\end{center}
\end{table}

\subsubsection{Two-phase optimization}
\label{subsec:twophaseOptimization}
As total optimization time is dominated by inner optimization time, reducing the number of outer iterations can substantially improve total optimization time. A strategy for speed-up is by using a two-phase optimization. In the early outer iterations (phase-I), full convergence of the beam weights is not important yet, because the $\delta$-factors will likely be updated to significantly different values. In later outer iterations (phase-II), more accurate optimizations are done. Therefore, in phase-I in the inner iterations we could use a) looser optimality tolerances (e.g., starting at \num{e-2} and increase their strictness to \num{e-4}), or use b) approximate Hessian schemes.

For case b, phase-I is done using the L-BFGS Hessian scheme with optimality tolerance \num{e-4}. We observed that this results in lower optimization times, while we keep obtaining significant $\delta$-factor updates in an improving direction. The convergence of phase-I is based on $L_2$-norm convergence with $\epsilon_s = 0.025$ for all structures with probabilistic objectives and constraints. Once phase-I has converged, we switch to phase-II (using the full Hessian implementation and optimality tolerance \num{e-5}). As the $\delta$-factors at the start of phase-II are a good estimation of the percentile already, the outer iterations in phase-II tend to converge faster than if phase-I was not used. To prevent that the optimization stops (because of percentile convergence) without converged beam weights, we only check for full percentile convergence in phase-II.

In the case that we use moving-average based convergence in phase-II, we start phase-II only if both phase-I has converged and the current iteration $k > W + \Delta K$. In such as case, it is beneficial to do a couple of faster phase-I optimizations, instead of starting phase-II (since the convergence criteria can not be calculated yet because of the involved lag).

If all probabilistic goals are achieved during optimization (or even at initialization), percentile convergence is not necessary to terminate the optimization. In such a case (checked only in phase-II, where exact Hessians are used), we only optimize until beam weight convergence. To confirm that all probabilistic objectives remain inactive and constraints remain satisfied, an additional optimization is done, starting from the undamped beam weights. If the probabilistic goals are still satisfied, we end the optimization, otherwise the next outer iteration is started.

The convergence criteria that were used for the probabilistic optimizations are summarized in Table \ref{tab:probConvSettings}. Optimizations for patient 1 and 2 were done using a single phase (1P, i.e., skipping phase-I), and reduced tolerances (RT) that become stricter over the course of the optimization. Patients 3, 4 and 5 were optimized using two phases (2P). Convergence in phase-II was based on moving average for all patients (and switching from phase-I to phase-II used the relative $L_2$-norm).

\subsubsection{Optimization times}
\begin{figure}[]
    \centering

    % First row
    \begin{subfigure}{0.43\textwidth}
        \centering
        \includegraphics[width=\linewidth]{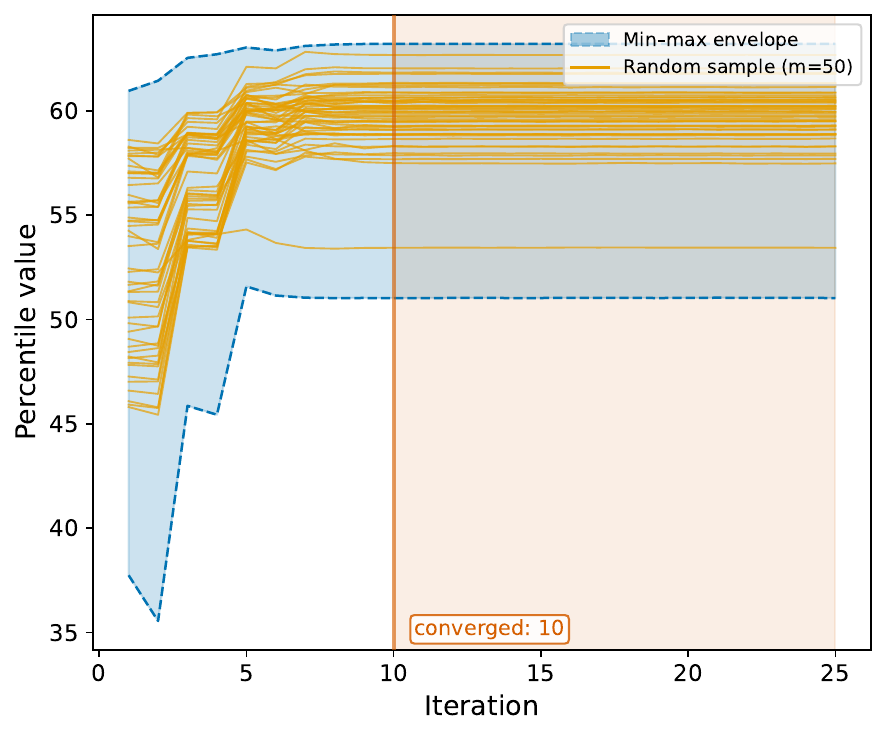}
        \caption{}
    \end{subfigure}
    %\hfill
    \begin{subfigure}{0.43\textwidth}
        \centering
        \includegraphics[width=\linewidth]{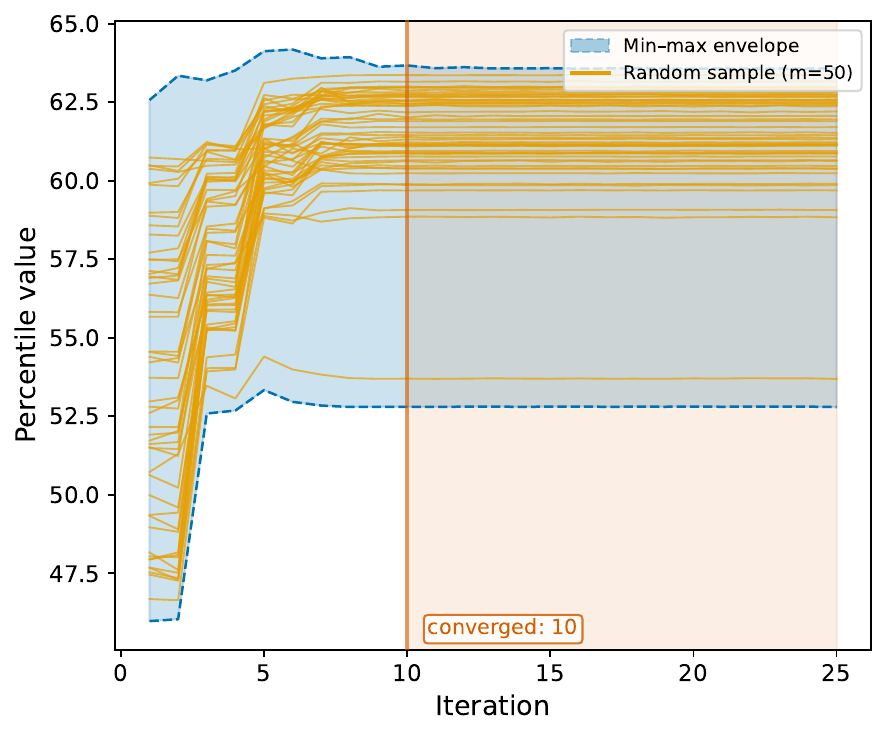}
        \caption{}
    \end{subfigure}
    %\hfill

    % Second row
    \begin{subfigure}{0.43\textwidth}
        \centering
        \includegraphics[width=\linewidth]{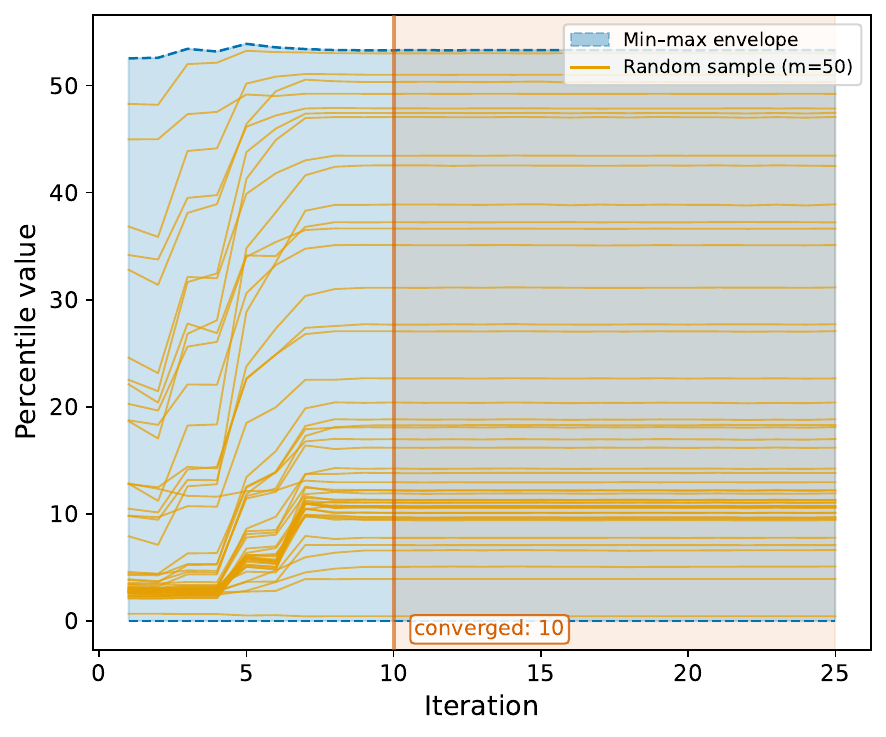}
        \caption{}
    \end{subfigure}
    \begin{subfigure}{0.43\textwidth}
        \centering
        \includegraphics[width=\linewidth]{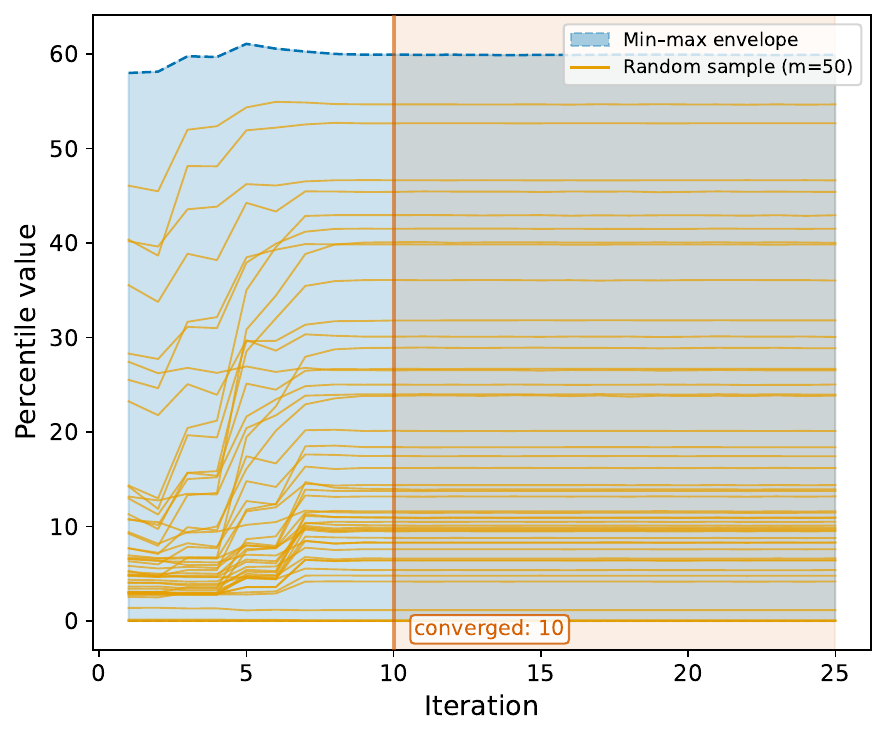}
        \caption{}
    \end{subfigure}

    \caption{Percentile convergence of patient 2 for the (a) 10th percentile and (b) 90th percentile of the CTV, for the 95th percentiles of the (c) Brainstem Core and (d) Brainstem Surface. The number of outer iterations is reduced by using different convergence settings. The minimum and maximum percentiles are shown as the envelope (blue) with 50 individual voxel dependencies.}
    \label{fig:convergenceBehaviour_Neuro10}
\end{figure}

Some convergence settings and tolerances in Table \ref{tab:probConvSettings} were chosen too conservative, leading to longer optimization times than necessary. An example of percentile convergence is shown in Figure \ref{fig:convergenceBehaviour_Neuro10} for patient 2. With the used convergence settings, 25 outer iterations (the maximum number of iterations that was set) were needed. However, by loosening the convergence settings listed in Table \ref{tab:probConvSettings}, the optimization converged after only 10 outer iterations. Optimization times reported in the manuscript are obtained by tuning the convergence criteria, done similarly as for patient 2. For patients 1 and 3, the number of outer iterations could be reduced by 2 iterations. \newline

\paragraph{\textbf{Optimization without warm-starting}}
To understand the impact of warm-starting the inner optimizations on the total optimization time, we performed additional optimizations for patient 4 and 5. For patient 4 the optimization was identical to the one in the manuscript. For patient 5, the maximum dose constraint of the brainstem core was changed from \qty{52.7}{\gray}RBE to \qty{51.7}{\gray}RBE. 

\begin{figure}[]
    \centering

    % First row
    \begin{subfigure}{0.48\textwidth}
        \centering
        \includegraphics[width=\linewidth]{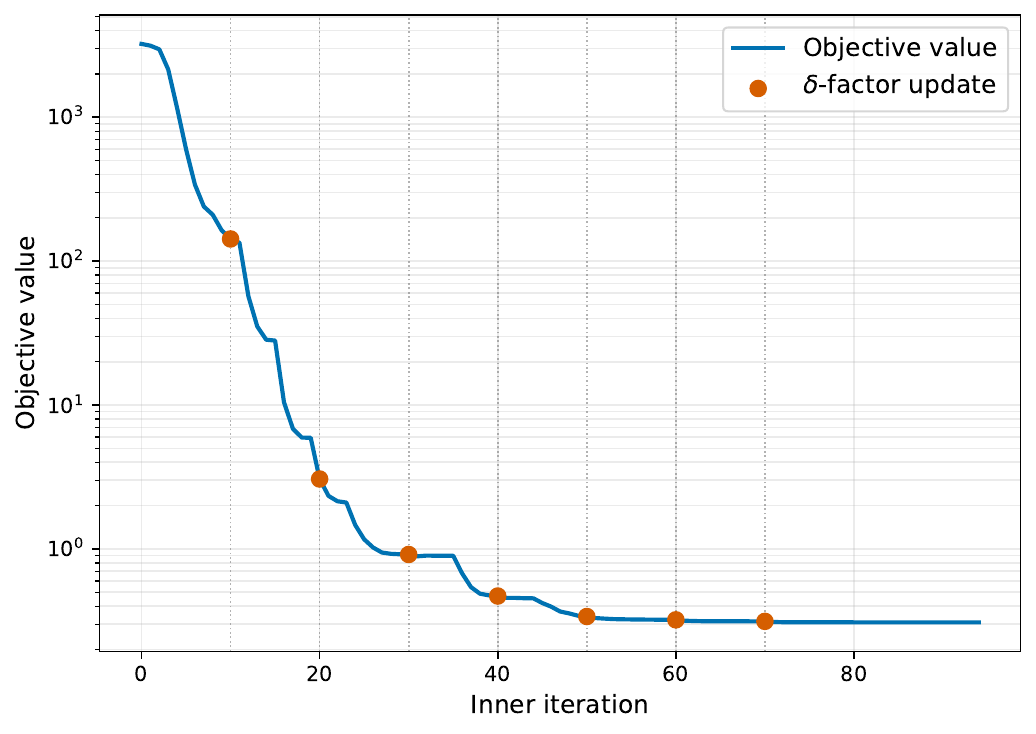}
        \caption{}
    \end{subfigure}
    \begin{subfigure}{0.48\textwidth}
        \centering
        \includegraphics[width=\linewidth]{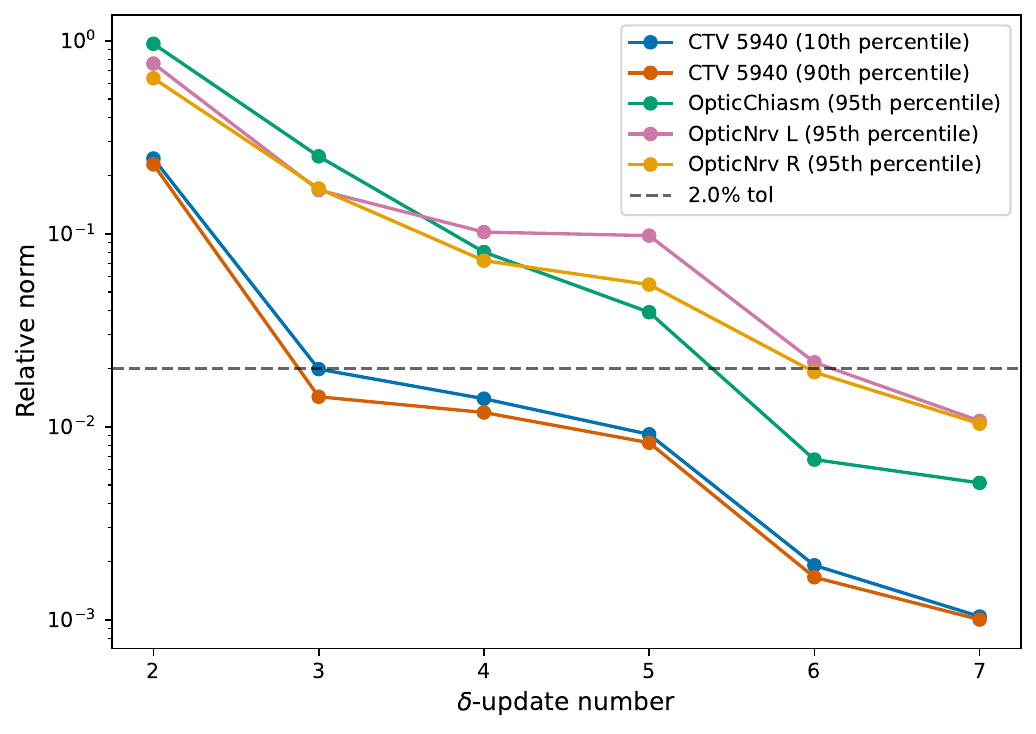}
        \caption{}
    \end{subfigure}

    % Second row
    \begin{subfigure}{0.48\textwidth}
        \centering
        \includegraphics[width=\linewidth]{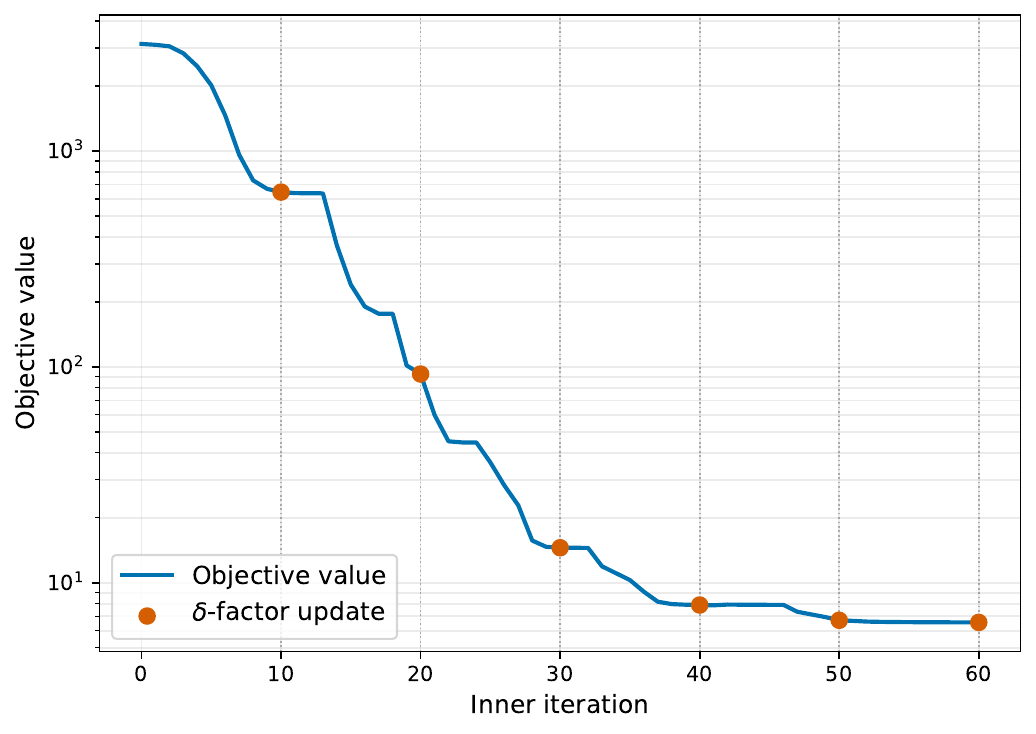}
        \caption{}
        \label{fig:noOuterLoop_convergence_c}
    \end{subfigure}
    \begin{subfigure}{0.48\textwidth}
        \centering
        \includegraphics[width=\linewidth]{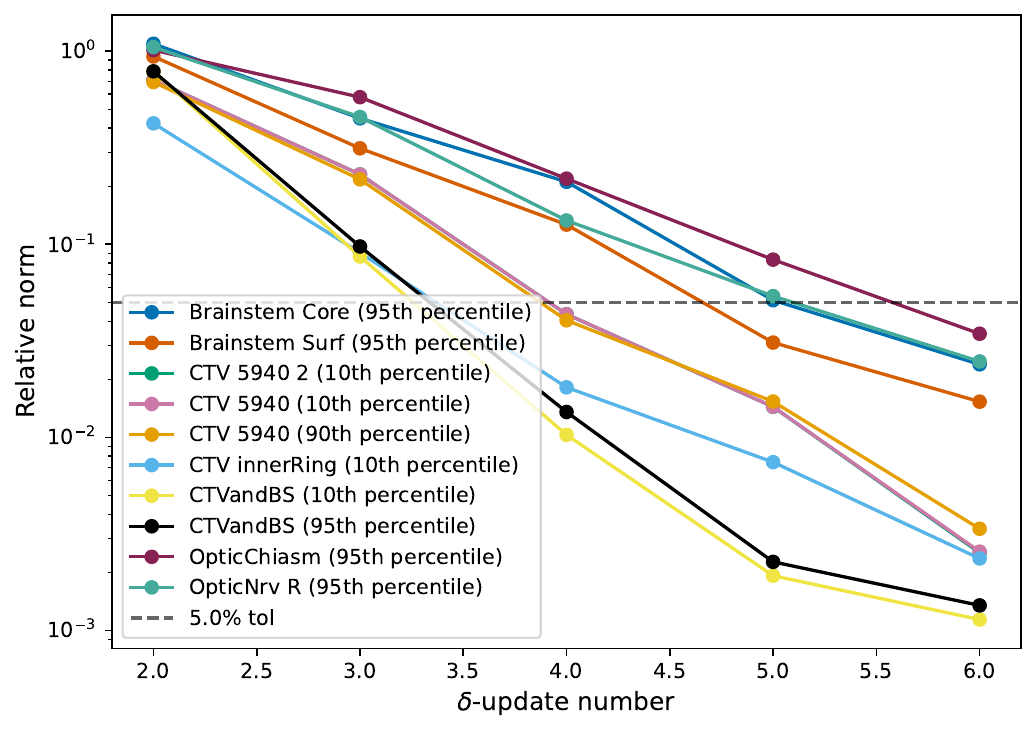}
        \caption{}
    \end{subfigure}
    
    \caption{Convergence results for patient 4 (top) and 5 (bottom), showing the objective value versus number of iterations (left: a, c) and the relative $L_2$-norm between successive percentile updates (right: b, d). The percentile updates are done every 10 iterations, see the orange dots in (a) and (c).}
    \label{fig:noOuterLoop_convergence}
\end{figure}

To eliminate warm-starting, we did not use an outer loop, but instead updated the $\delta$-factors within a single (inner) optimization. We update the $\delta$-factors every 10 (inner) iterations until the relative $L_2$-norm of the voxel dose percentiles between successive percentile updates was below $\epsilon_s, \, \forall s \in S_{prob}$, i.e., 
\begin{equation}
    \Delta \bi{d}_s^{\alpha \%}(k) = \frac{\Vert \bi{d}^{\alpha \%}_{k,s} - \bi{d}^{\alpha \%}_{k-10,s} \Vert}{\Vert \bi{d}^{\alpha \%}_{k-10,s} \Vert + \num{e-10} } < \epsilon_s,
\end{equation}
for $k =\{ 10, 20, 30, \ldots \}$. Once $\Delta \bi{d}_s^{\alpha \%}(k) < \epsilon_s, \, \forall s \in S_{prob}$, the optimization continues until beam weight convergence is reached using the current $\delta$-factors. The convergence results for the probabilistic optimization without outer loop are shown in Figure \ref{fig:noOuterLoop_convergence}, where the iteration number corresponding to $\delta$-factor updates is highlighted by orange dots. Damping of $\delta$-factors with damping factor $\kappa$ was used in this optimization for stability, using
\begin{equation}
    \boldsymbol{\delta}_{\text{next}}^s = (1 - \kappa) \cdot \boldsymbol{\delta}_{\text{previous}}^s + \kappa \cdot \boldsymbol{\delta}_{\text{current}}^s,
\end{equation}
where the next $\delta$-factors are obtained by damping between the current ($\boldsymbol{\delta}_{\text{current}}^s$) and previous ($\boldsymbol{\delta}_{\text{previous}}^s$) factors. 

\begin{table}
\caption{\label{tab:noWarmStarting}Optimization and convergence settings for the probabilistic optimizations without warm starting (i.e., outer loop) of patients 4 and 5.}
\begin{center}
\begin{tabular}{@{}lcc}
\toprule
 & Patient 4 & Patient 5 \\[0.5ex]
\toprule
Damping factor ($\kappa$)                                                        & 0.3          & 0.1            \\
Convergence tolerance ($\epsilon_s, \, \forall s \in S_{\mathrm{prob}}$)         & 0.02         & 0.05           \\
Optimality tolerance                                                             & \num{1e-5}   & \num{4.5e-5}   \\
\midrule
Total optimization time (original) (\unit{\hour})                                & 20.3         & 141            \\
Total optimization time (no warm-start) (\unit{\hour})                           & 9.8          & 8.6            \\
\bottomrule
\end{tabular}
\end{center}
\end{table}

The optimization and convergence settings for the two probabilistic optimizations without warm-starting are listed in Table \ref{tab:noWarmStarting}. For patient 4, plan quality was similar compared to the optimization presented in the manuscript (OAR-related DVH-metrics where identical within within 1.7\%, only considering metrics above $\qty{3}{\gray}\text{RBE}$). Because for patient 5, the wish-list was slightly different from the optimization in presented in the manuscript, comparing plan quality would be less meaningful. Regardless, this analysis was meant to evaluate the computational impact of warm-starting.

For patient 4 and 5, the total optimization time could be reduced by 88\% and 94\%, respectively. The plan corresponding to patient 5 converged within an optimality tolerance of \num{4.5e-5} (instead of \num{e-5}). A tolerance of \num{e-5} could not be reached, because the optimization stopped due to a small step size (with step size tolerance \num{1e-25}), just after the $\delta$-factors were updated (at inner iteration 60, see Figure \ref{fig:noOuterLoop_convergence_c}). However, this does not affect the conclusion of the analysis, namely that significant time reduction could be achieved by removing warm-starting from the probabilistic planning approach.

\end{document}